\documentclass[12 pt]{article}
\pdfoutput=1

\usepackage{fullpage}
\usepackage{amscd}
\usepackage{amsmath}
\usepackage{amssymb}
\usepackage{amsthm}
\usepackage{epsf}
\usepackage{epsf}
\usepackage{graphicx}
\usepackage{latexsym}
\usepackage{verbatim}
\usepackage{tikz}
\usetikzlibrary{backgrounds,shapes.geometric,fit,matrix,arrows}
\input epsf.tex

\newtheorem{theorem}{Theorem}[section]
\newtheorem{lemma}[theorem]{Lemma}
\newtheorem{corollary}[theorem]{Corollary}
\newtheorem{proposition}[theorem]{Proposition}
\newtheorem*{nashthm}{Nash's Theorem}
\newtheorem*{minimaxthm}{Minimax Theorem}

\theoremstyle{definition}

\theoremstyle{definition}

\theoremstyle{remark}
\newtheorem{example}[theorem]{Example}

\theoremstyle{remark}

\theoremstyle{definition} % Doesn't agree with IEEE .cls file
\newtheorem{definition}[theorem]{Definition}

\newcounter{tempthm}
\newcounter{tempsec}

\newcommand{\savecounter}[1]{\newcounter{thmcounter#1}
\setcounter{thmcounter#1}{\value{theorem}}
\newcounter{seccounter#1}
\setcounter{seccounter#1}{\value{section}}}

\newcommand{\usesavedcounter}[1]{\setcounter{tempthm}{\value{theorem}}
\setcounter{theorem}{\value{thmcounter#1}}
\setcounter{tempsec}{\value{section}}
\setcounter{section}{\value{seccounter#1}}}

\newcommand{\restorecounter}{\setcounter{theorem}{\value{tempthm}}
\setcounter{section}{\value{tempsec}}}

\newcommand{\Nm}{\mathbb{N}}

\newcommand{\Rm}{\mathbb{R}}

\newcommand{\Zm}{\mathbb{Z}}

\newcommand{\mpowm}[2]{\Pi^{#1} #2}
\newcommand{\mpow}[1]{\mpowm{m}{#1}}
\newcommand{\compowm}[2]{\Xi^{#1} #2}
\newcommand{\compow}[1]{\compowm{m}{#1}}

\newcommand{\Sym}{\text{Sym}}

\newcommand{\shortv}[1]{}

\DeclareMathOperator{\conv}{conv}
\DeclareMathOperator{\clconv}{\overline{conv}}

\DeclareMathOperator{\prob}{\mathbb{P}}
\DeclareMathOperator{\expect}{\mathbb{E}}

\DeclareMathOperator{\ce}{CE}
\DeclareMathOperator{\nash}{NE}
\DeclareMathOperator{\xe}{XE}

\DeclareMathOperator{\maximin}{Mm}

\DeclareMathOperator{\ave}{ave}
\DeclareMathOperator{\proj}{proj}

\newcommand{\eqdef}{:=}

\providecommand{\abs}[1]{\lvert#1\rvert}

\title{\LARGE \bf
A new proof of Nash's Theorem via exchangeable equilibria
}

\author{Noah D. Stein, Pablo A. Parrilo, and Asuman Ozdaglar% <-this % stops a space
\thanks{Department of Electrical Engineering,
        Massachusetts Institute of Technology: Cambridge, MA 02139.
        {\tt\small nstein@mit.edu}, {\tt\small parrilo@mit.edu}, and {\tt\small asuman@mit.edu}.}
      }

\begin{document}

\maketitle

\thispagestyle{empty}

\pagestyle{plain}

\begin{abstract}
We give a novel proof of the existence of Nash equilibria in all finite games without using fixed point theorems or path following arguments.  Our approach relies on a new notion intermediate between Nash and correlated equilibria called exchangeable equilibria, which are correlated equilibria with certain symmetry and factorization properties.  We prove these exist by a duality argument, using Hart and Schmeidler's proof of correlated equilibrium existence as a first step.

In an appropriate limit exchangeable equilibria converge to the convex hull of Nash equilibria, proving that these exist as well.  Exchangeable equilibria are defined in terms of symmetries of the game, so this method automatically proves the stronger statement that a symmetric game has a symmetric Nash equilibrium.  The case without symmetries follows by a symmetrization argument.
\end{abstract}

\section{Introduction}
Nash's Theorem is one of the most fundamental results in game theory and states that any finite game has a Nash equilibrium in mixed strategies.  Despite its importance, the authors of the present paper know of only three essentially different proofs.  The first and most common way to prove Nash's Theorem is by applying a fixed point theorem, usually Brouwer's or Kakutani's, to a map whose fixed points are easily shown to be Nash equilibria.  The fixed point theorem is usually proven combinatorially, say by Sperner's Lemma \cite{sperner} or Gale's argument using the game of hex \cite{gale:hexbrouwer}, or with (co-)homology theory, a suite of powerful but less elementary tools from algebraic topology \cite{hatcher:at}.

The second proof of Nash's Theorem (historically) is algorithmic and consists of showing that the Lemke-Howson path-following algorithm terminates at a Nash equilibrium \cite{lh:epbg}.  In fact this is not so different from the fixed point proof, because Sperner's Lemma is also proven by a path-following argument.

The third proof of Nash's theorem is due to Kohlberg and Mertens and is topological \cite{km:osse}.  The idea is to simultaneously consider the set of all games of a given size and the set of all $(\text{game},\text{equilibrium})$ pairs.  Under appropriate compactifications both of these sets become spheres of the same dimension.  One then shows that the mapping sending an equilibrium to the corresponding game is homotopic to the identity map on the sphere.  A (co-)homological or degree-theoretic argument shows that such a map must be onto \cite{hatcher:at}.  Technically speaking this step is almost identical to the proof of Brouwer's Fixed Point Theorem, so the third proof is closely related to the first.

All three proofs have provided unique insights into the structure of Nash equilibria and it is our hope that a different proof, which uses neither fixed point theorems, nor path-following arguments, nor any algebro-topological tools, will provide further insights.

Hart and Schmeidler \cite{hs:ece} have proven the weaker result that correlated equilibria exist by a clever application of the Minimax Theorem, summarized in Section \ref{sec:hs}.  For games endowed with a group action, a simple averaging argument then proves that a symmetric correlated equilibrium exists (Proposition \ref{prop:symcorrexist}).  We show that for such games Hart and Schmeidler's proof can be strengthened to produce correlated equilibria with additional symmetry and factorization properties, which we call \emph{exchangeable equilibria} (Theorem \ref{thm:symexeqexist}).

To illustrate this idea, consider the case of $k\times k$ symmetric bimatrix games (two-player games fixed under the operation of swapping the players).  Let $X = \{xx^T \mid x\in\Rm^{k\times 1}_{\geq 0}\}$.  Then we have
\[
\stackrel{\text{Nash}}{CE\cap X}\ \subseteq\ \stackrel{\text{convex hull of Nash}}{\conv(CE\cap X)}\ \subseteq\ \stackrel{\text{exchangeable}}{CE\cap\conv(X)}\ \subseteq\ \stackrel{\text{correlated}}{CE},
\]
where each type of (symmetric) equilibrium is defined by the set written below it.  This definition shows that the exchangeable equilibria are a natural mathematical object.  For examples and game-theoretic interpretations of exchangeable equilibria, see the companion paper \cite{sop:xe1}.

In Section~\ref{sec:exeq} we extend the definition of exchangeable equilibria games to $n$-player games with arbitrary symmetry groups.  The preceding discussion shows that the set of exchangeable equilibria is convex, compact, contained in the set of symmetric correlated equilibria, and contains the convex hull of the set of symmetric Nash equilibria.  One can show that these containments can all be strict \cite{sop:xe1}, so proving existence of exchangeable equilibria is a step in the right direction, but does not immediately prove existence of Nash equilibria.

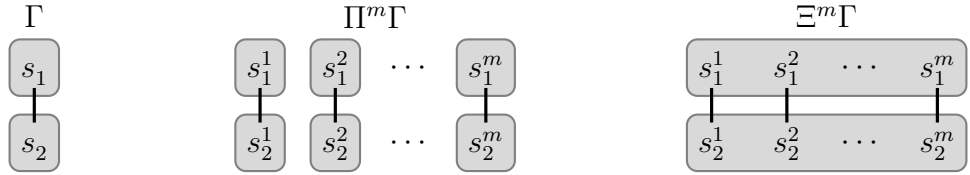
\begin{figure}[htbp]
\begin{center}
\begin{tikzpicture}
[outline/.style={draw=gray},
 strat/.style={inner sep=1pt,text height=10pt},%{rounded corners,outline,fill=gray!60,thick,inner sep=2pt},
 contract/.style={rounded corners,outline,fill=gray!30,thick,inner sep=3pt}]
 \node (Gamma2) at (0,0) [strat] {$s_2^{\ }$};
 \node (Gamma1) at (0,1) [strat] {$s_1^{\ }$};
 
 \node (Pi21) at (3,0) [strat] {$s_2^1$};
 \node (Pi22) at (4,0) [strat] {$s_2^2$};
 \node at (5,0) {$\cdots$};
 \node (Pi2m) at (6,0) [strat] {$s_2^m$};
 \node (Pi11) at (3,1) [strat] {$s_1^1$};
 \node (Pi12) at (4,1) [strat] {$s_1^2$};
 \node at (5,1) {$\cdots$};
 \node (Pi1m) at (6,1) [strat] {$s_1^m$};
 
 \node (Xi21) at (9,0) [strat] {$s_2^1$};
 \node (Xi22) at (10,0) [strat] {$s_2^2$};
 \node at (11,0) {$\cdots$};
 \node (Xi2m) at (12,0) [strat] {$s_2^m$};
 \node (Xi11) at (9,1) [strat] {$s_1^1$};
 \node (Xi12) at (10,1) [strat] {$s_1^2$};
 \node at (11,1) {$\cdots$};
 \node (Xi1m) at (12,1) [strat] {$s_1^m$};
 
  \draw [-,very thick] (Gamma1) -- (Gamma2); 
 
 \draw [-,very thick] (Pi11) -- (Pi21); 
 \draw [-,very thick] (Pi12) -- (Pi22); 
 \draw [-,very thick] (Pi1m) -- (Pi2m);
 
 \draw [-,very thick] (Xi11) -- (Xi21); 
 \draw [-,very thick] (Xi12) -- (Xi22); 
 \draw [-,very thick] (Xi1m) -- (Xi2m);
 \begin{pgfonlayer}{background}
\node (Gamma1prime) [contract,fit=(Gamma1),label=above:$\Gamma$] {};
\node (Gamma2prime) [contract,fit=(Gamma2)] {}; 

%\draw [-,very thick] (Gamma1prime) -- (Gamma2prime);

\node[inner sep=3pt,fit=(Pi11) (Pi1m),label=above:$\mpow{\Gamma}$] {};
\node[contract,fit=(Pi11)] {};
\node[contract,fit=(Pi12)] {};
\node[contract,fit=(Pi1m)] {};
\node[contract,fit=(Pi21)] {};
\node[contract,fit=(Pi22)] {};
\node[contract,fit=(Pi2m)] {};
 
\node[contract,fit=(Xi11) (Xi1m),label=above:$\compow{\Gamma}$] {};
\node[contract,fit=(Xi21) (Xi2m)] {}; 
\end{pgfonlayer} 
\end{tikzpicture}
\caption{Representing a $2$-player game $\Gamma$ with players choosing strategies $s_1$ and $s_2$ as drawn on the left, the powers $\mpow{\Gamma}$ and $\compow{\Gamma}$ are formed as shown.  A shaded box represents actions controlled by a single player, whose utility is given by the sum over all interactions.}
\label{fig:powers}
\end{center}
\end{figure}

However, we can use the same techniques to prove existence of exchangeable equilibria with successively stronger symmetry properties as follows.  For a fixed $n$-player game $\Gamma$ and number $m\in\Nm$, we define two new games $\mpow{\Gamma}$ and $\compow{\Gamma}$, which we call \emph{$m^\text{th}$ powers of $\Gamma$}.  These are larger games in which $m$ copies of $\Gamma$ are played simultaneously.  The difference between the two powers is that $\mpow{\Gamma}$ has a different group of players for each copy, so $mn$ players total, whereas $\compow{\Gamma}$ has just one group of $n$ players playing all $m$ copies, but perhaps choosing different actions in each copy (Figure \ref{fig:powers}).

There is a natural marginalization map sending exchangeable equilibria of either of these powers to exchangeable equilibria of $\Gamma$.  In fact, any exchangeable equilibrium of $\Gamma$ can be lifted to an exchangeable equilibrium of either power, and for a symmetric Nash equilibrium we can take these two lifts to be the same.  However, for a general exchangeable equilibrium the two lifts need not be compatible, so it is natural to consider the intersection $\xe^m(\Gamma)\eqdef\xe(\mpow{\Gamma})\cap\xe(\compow{\Gamma})$ of the sets of exchangeable equilibria of the two powers.  We call these \emph{order $m$ exchangeable equilibria of $\Gamma$} and prove they exist using a similar Minimax argument (Theorem~\ref{thm:symordermexeqexist}).

Under appropriate identifications these sets turn out to be nested as $m$ grows and, being convex, they all contain the convex hull of the symmetric Nash equilibria:
\begin{equation*}
\nash(\Gamma)\subseteq\conv(\nash(\Gamma))\subseteq\ldots\subseteq\xe^3(\Gamma)\subseteq\xe^2(\Gamma)\subseteq\xe^1(\Gamma)=\xe(\Gamma)\subseteq\ce(\Gamma).
\end{equation*}
Assuming $\Gamma$ has a rich enough symmetry group (e.g.\ if it is a symmetric bimatrix game or, more generally, if it satisfies a condition we call \emph{player transitivity}), then as $m$ goes to infinity these converge to symmetric correlated equilibria in which the outcome of the correlating device is common knowledge; such correlated equilibria are known to be mixtures of symmetric Nash equilibria (Theorems~\ref{thm:symorderinfexeqexist} and \ref{thm:orderinfexeqchar}).  In particular, this proves that symmetric Nash equilibria exist in games with rich enough symmetry groups.

Note that symmetry is fundamental in this argument.  For example, if we had begun with a general bimatrix game and let $X = \{xy^T \mid x,y\in\Rm^{k\times 1}_{\geq 0}\}$ we would have had $\conv(X) = \Rm^{k\times k}_{\geq 0}$, so the exchangeable equilibria (even the order $m$ exchangeable equilibria) would have been exactly the correlated equilibria and we would not have strengthened the equilibrium notion at all.  However, there are several ways of turning general games into symmetric games \cite{gkt:osg} and applying such a procedure proves existence of Nash equilibria in all games (Section \ref{sec:finishproof}).

Up to the step of taking $m$ to infinity, all the steps of our proof are computationally effective.  Papadimitriou has shown how to apply the ellipsoid algorithm to Hart and Schmeidler's proof to efficiently compute a correlated equilibrium of a large game \cite{p:ccempg}.  The same technique applied to our proof allows one to compute an exchangeable equilibrium (or an order $m$ exchangeable equilibrium for fixed $m$) in polynomial time, even though the set of these is not polyhedral.  Computing these is interesting in its own right \cite{sop:xe1} and may be useful for computing approximate Nash equilibria.  However, computation is not the focus of this paper and we leave a detailed investigation of these ideas for future work.

The remainder of the paper is organized as follows.  We begin with background material in Section~\ref{sec:background}.  We cover the definitions of games and equilibria, give an overview of Hart and Schmeidler's proof of the existence of correlated equilibria so we can modify it later, and introduce group actions.  In Section~\ref{sec:exeq} we introduce exchangeable equilibria and prove existence of these for games under arbitrary group actions.  We do the same for order $m$ exchangeable equilibria in Section~\ref{sec:higherexeq}, introducing powers of games along the way.  We complete the argument in Section~\ref{sec:nash} by showing that the order $m$ exchangeable equilibria converge to mixtures of symmetric Nash equilibria in games with a player transitive symmetry group, and then showing that we can symmetrize any game to make this condition hold.  Section~\ref{sec:conclusions} concludes and gives directions for future work.

\section{Background}
\label{sec:background}
This section is divided into three parts.  In the first part we lay out the basic definitions of finite games as well as Nash and correlated equilibria to fix notation.  We assume the reader is familiar with these concepts and do not attempt to motivate them.  The second part reviews Hart and Schmeidler's proof of the existence of correlated equilibria \cite{hs:ece}, preparing for similar arguments in this paper.  The third part covers symmetries of games.

The concept of a symmetry of a game extends back at least to Nash's paper \cite{nash:ncg}.  Symmetries are fundamental to the present paper, so we spend more time on these and give some examples.  Although we use the language of group theory to discuss symmetries, it is worth noting that we do not use any but the most basic theorems from group theory (e.g., the fact that for any $h$ in a group $G$, the maps $g\mapsto gh$ and $g\mapsto hg$ are bijections from $G$ to $G$).  Everything in this section is standard except for the notions of a \emph{good reply}, a \emph{good set}, a \emph{player-trivial symmetry group}, a \emph{player-transitive symmetry group}, and the remarks following the statement of Nash's Theorem.

\subsection{Games and equilibria}
\label{sec:games}
\begin{definition}
A \textbf{(finite) game} has a finite set $I$ of $n\geq 2$ players, each with a finite set $C_i$ of at least two \textbf{strategies} (also called \textbf{pure strategies}) and a \textbf{utility function} $u_i: C\to\Rm$, where $C = \prod C_i$.  A game is \textbf{zero-sum} if it has two players, called the \textbf{maximizer} (denoted $M$) and the \textbf{minimizer} (denoted $m$), and satisfies $u_M + u_m \equiv 0$.
\end{definition}

For elements of $C_i$ we use Roman letters subscripted with the player's identity, such as $s_i$ and $t_i$.  We will typically use the unsubscripted letter $s$ to denote a strategy profile (a choice of strategy for each player).  For a choice of a strategy for all players except $i$ we use the symbol $s_{-i}$.  To denote the set of Borel probability distributions on a space $X$ we write $\Delta(X)$.  For much of the paper $X$ will be finite so we can view $\Delta(X)$ as a simplex in the finite-dimensional vector space $\Rm^X$ of real-valued functions on $X$.  For $x\in X$ the probability distribution which assigns unit mass to $x$ will be written $\delta_x\in\Delta(X)$.

\begin{definition}
A \textbf{mixed strategy} for player $i$ in a game $\Gamma$ is a probability distribution over his pure strategy set $C_i$, and the set of mixed strategies for player $i$ is $\Delta(C_i)$.  The set of \textbf{mixed strategy profiles} (also called \textbf{independent} or \textbf{product distributions}) will be denoted $\Delta^\Pi(\Gamma) \eqdef \prod_i \Delta(C_i)$.
\end{definition}

For independent distributions it is important that we write $\Delta^\Pi(\Gamma)$ rather than $\Delta^\Pi(C)$, because $\Gamma$ specifies how $C$ is to be thought of as a product.  For example, the set $S\times S\times S$ could be viewed as a product of three copies of $S$, or a product of $S$ with $S\times S$, and these lead to different notions of an independent distribution -- one is a product of three terms and one is a product of two terms.  This distinction will be particularly important when we define powers of games in Section~\ref{sec:higherexeq}.

To make the notation fit together we will write $\Delta(\Gamma)$ for $\Delta(C)$.  We may then view $\Delta^\Pi(\Gamma)$ as the (nonconvex) subset of $\Delta(\Gamma)$ consisting of product distributions or as a convex subset of $\Rm^{\sqcup_i C_i}$.  The former view will be natural when we define exchangeable equilibria, which live in $\Delta(\Gamma)$, as convex combinations of product distributions.  The latter will be useful when looking for product distributions which are fixed by a group action (see the proof of Lemma~\ref{lem:exeqgood}); such fixed distributions are easy to find with a convex setup (Proposition~\ref{prop:average}).  Which of these views we are using will be clear from context if not explicitly specified.

As usual we extend the domain of $u_i$ from $C$ to $\Delta(\Gamma)$ by linearity, defining $u_i(\pi) = \sum_{s\in C} u_i(s)\pi(s)$.  Having done so we can define equilibria.

\begin{definition}
A \textbf{Nash equilibrium} is an $n$-tuple $(\rho_1,\ldots,\rho_n)\in\Delta^\Pi(\Gamma) = \prod_i \Delta(C_i)$ of mixed strategies, one for each player, such that $u_i(s_i,\rho_{-i}) \leq u_i(\rho_i,\rho_{-i})$ for all strategies $s_i\in C_i$ and all players $i$.  The set of Nash equilibria of a game $\Gamma$ is denoted $\nash(\Gamma)$.
\end{definition}

\begin{definition}
\label{def:correq}
A \textbf{correlated equilibrium} is a joint distribution $\pi\in\Delta(\Gamma)$ such that $\sum_{s_{-i}\in C_{-i}}\left[u_i(t_i,s_{-i}) - u_i(s)\right]\pi(s)\leq 0$ for all strategies $s_i,t_i\in C_i$ and all players $i$.  The set of correlated equilibria of a game $\Gamma$ is denoted $\ce(\Gamma)$.
\end{definition}

The following alternative characterization of correlated equilibria is standard and we omit its proof.  Suppose $(X_1,\ldots, X_n)$ is a random vector taking values in $C$.  We think of $X_i$ as a (random) strategy recommended to player $i$.  Given this information, player $i$ can form his conditional beliefs $\prob(X_{-i} \mid X_i)$ about the recommendations to the other players given his own recommendation.  That is to say, $\prob(X_{-i} \mid X_i)$ is a random variable taking values in $\Delta(C_{-i})$ which is a function of $X_i$.  One can then define the event
\begin{equation*}
\{\text{the pure strategy }X_i\text{ is a best response to the distribution }\prob(X_{-i}\mid X_i)\text{ for all }i\}.
\end{equation*}
The distribution of $(X_1,\ldots,X_n)$ is a correlated equilibrium if and only if this event happens almost surely.  More succinctly:

\begin{proposition}
\label{prop:correqchar}
Let $(X_1,\ldots,X_n)$ be a random vector taking values in $C$ distributed according to $\pi\in\Delta(\Gamma)$.  Then $\pi$ is a correlated equilibrium if and only if $X_i$ is a best response to $\prob(X_{-i}\mid X_i)$ almost surely for all $i$.
\end{proposition}

Nash equilibria correspond exactly to the correlated equilibria which are product distributions, so viewing $\Delta^\Pi(\Gamma)$ as a subset of $\Delta(\Gamma)$ we can write $\nash(\Gamma) = \ce(\Gamma)\cap\Delta^\Pi(\Gamma)$.  We introduce the existence theorems for correlated and Nash equilibria in Sections~\ref{sec:hs} and~\ref{sec:sym}.

We need the Minimax Theorem at this point to define the value of a zero-sum game.  It also plays an important role in our proof of Nash's Theorem.  %The Minimax Theorem is perhaps the only result in game theory which could be said to be more fundamental than Nash's Theorem.
An elementary proof can be given using the separating hyperplane theorem \cite{bno:convex}.

\begin{minimaxthm}
Let $U$ and $V$ be finite-dimensional vector spaces with compact convex subsets $K\subset U$ and $L\subset V$.  Let $\Phi: U\times V\to \Rm$ be a bilinear map.  Then
\begin{equation*}
\sup_{x\in K}\inf_{y\in L} \Phi(x,y) = \inf_{y\in L}\sup_{x\in K} \Phi(x,y),
\end{equation*}
and the optima are attained.
\end{minimaxthm}

\begin{definition}
Given a zero-sum game $\Gamma$, we can apply this theorem with $K = \Delta(C_M)$, $L=\Delta(C_m)$, and $\Phi = u_M$.  The common value of these two optimization problems is called the \textbf{value} of the game and denoted $v(\Gamma)$.  Maximizers on the left hand side are called \textbf{maximin strategies} and the set of such is denoted $\maximin(\Gamma)\subseteq\Delta(C_M)$.
\end{definition}

We now introduce the notion of a \emph{good reply} in a zero-sum game.  This is not a standard definition, but it will simplify the statements of several arguments below.  The name is meant to be evocative of the term \emph{best reply}: while a best reply is one which maximizes one's payoff, a good reply is merely one which returns a ``good'' payoff: at least the value of the game\footnote{\emph{Good sets} are similar in spirit to Voorneveld's \emph{prep sets} \cite{v:p}, but tailored to zero-sum games.}.

\begin{definition}
\label{def:good}
In a zero-sum game $\Gamma$, we say that a strategy $\sigma\in\Delta(C_M)$ for the maximizer is a \textbf{good reply to $\theta\in\Delta(C_m)$} if $u_M(\sigma,\theta)\geq v(\Gamma)$.  We say that a set $\Sigma\subseteq\Delta(C_M)$ of strategies is \textbf{good against the set $\Theta\subseteq\Delta(C_m)$} if for all $\theta\in\Theta$ there is a $\sigma\in\Sigma$ which is a good reply to $\theta$.  If $\Sigma$ is good against $\Delta(C_m)$ we say that $\Sigma$ is \textbf{good}.
\end{definition}

The main result about good sets is:

\begin{proposition}
\label{prop:minimax}
If $\Gamma$ is a zero-sum game and $\Sigma\subseteq\Delta(C_M)$ is good, then $\Gamma$ has a maximin strategy in $\clconv(\Sigma)$, i.e., $\clconv(\Sigma)\cap\maximin(\Gamma)\neq\emptyset$.
\end{proposition}
\begin{proof}
Apply the Minimax Theorem with $K = \clconv(\Sigma)$ and $L = \Delta(C_m)$.
\end{proof}

It is worth noting that in general a good set need not include a maximin strategy.  For example, in any zero-sum game the set $C_M\subsetneq\Delta(C_M)$ is a good set, but some zero-sum games such as matching pennies only have mixed maximin strategies, i.e. $C_M\cap\maximin(\Gamma)=\emptyset$.

The notion of payoff equivalence is a standard way to turn structural information about a game into structural information about equilibria.  The proofs of both propositions below are immediate.

\begin{definition}
Two mixed strategies $\sigma_i,\tau_i\in\Delta(C_i)$ are said to be \textbf{payoff equivalent} if $u_j(\sigma_i,s_{-i}) = u_j(\tau_i,s_{-i})$ for all $s_{-i}\in C_{-i}$ and all players $j$.
\end{definition}

\begin{proposition}
\label{prop:stratequivgood}
If $\Gamma$ is a zero-sum game, $\Sigma$ is good, and each $\sigma\in\Sigma$ is payoff equivalent to some $\sigma'\in\Sigma'$, then $\Sigma'$ is good.
\end{proposition}

\begin{proposition}
\label{prop:stratequivnash}
If $\sigma_i$ is payoff equivalent to $\tau_i$ for all $i$, then $(\sigma_1,\ldots,\sigma_n)$ is a Nash equilibrium if and only if $(\tau_1,\ldots,\tau_n)$ is a Nash equilibrium.
\end{proposition}

\subsection{Hart and Schmeidler's proof}
\label{sec:hs}
In this section we recall the structure of Hart and Schmeidler's proof of the existence of correlated equilibria based on the Minimax Theorem \cite{hs:ece}.  The goal of this is to frame their argument in a way which will allow us to extend it, redoing as little as possible of the work they have done.  We will use similar arguments to prove Theorems~\ref{thm:symexeqexist} and~\ref{thm:symordermexeqexist}.

Hart and Schmeidler's argument begins by associating with a game $\Gamma$ a new zero-sum game $\Gamma^0$ and interpreting correlated equilibria of $\Gamma$ as maximin strategies of this new game.  In $\Gamma^0$ the maximizer plays the roles of all the players in $\Gamma$ simultaneously and the minimizer tries to find a profitable unilateral deviation from the strategy profile selected by the maximizer.

\begin{definition}
\label{def:gamma0}
Given any game $\Gamma$, define a two-player zero-sum game $\Gamma^0$ with $C_M^0 \eqdef C$, $C_m^0 \eqdef \bigsqcup_i C_i\times C_i$, and utilities
\begin{equation*}
u_M^0(s,(r_i,t_i)) = -u_m^0(s,(r_i,t_i)) \eqdef \begin{cases}u_i(s) - u_i(t_i,s_{-i})&\text{ if }r_i = s_i, \\ 0 & \text{otherwise}.\end{cases}
\end{equation*}
\end{definition}

\begin{proposition}
\label{prop:gamma0}
Let $\Gamma$ be any game.  For any player $i$ in $\Gamma$, $r_i\in C_i$, and $s\in C$ we have $u_M^0(s,(r_i,r_i))=0$, so we can bound the value of $\Gamma^0$ by $v(\Gamma^0)\leq 0$.  A mixed strategy $\sigma\in\Delta(C^0_M) = \Delta(C)$ for the maximizer in $\Gamma^0$ satisfies $u_M^0(\sigma,(r_i,t_i))\geq 0$ for all $(r_i,t_i)\in C_m^0$ if and only if $\sigma\in\ce(\Gamma)$.  Therefore, if $v(\Gamma^0)=0$ then $\maximin(\Gamma^0) = \ce(\Gamma)$.
\end{proposition}

\begin{proof}Immediate from the definitions.
\end{proof}

To prove $v(\Gamma^0) = 0$, and hence the existence of correlated equilibria (Theorem~\ref{thm:hs}), we must show that for any $y\in\Delta(C_m^0)$ there is a $\pi\in\Delta(C_M^0)$ such that $u_M(\pi,y)\geq 0$.  Hart and Schmeidler actually show that there exists such a $\pi$ with some extra structure, which we summarize in Lemma~\ref{lem:hs}.  We will exploit this extra structure below to prove stronger statements in a similar spirit: Lemmas~\ref{lem:exeqgood} and \ref{lem:ordermexeqgood}.  These in turn allow us to prove the existence of exchangeable equilibria (Theorem~\ref{thm:symexeqexist}) and order $m$ exchangeable equilibria (Theorem~\ref{thm:symordermexeqexist}).

Given a $y = (y_1,\ldots,y_n)\in\Delta(C_m^0)$, $y_i\in\Rm^{C_i\times C_i}$, a good reply $\pi$ can be constructed in terms of certain auxiliary games $\gamma(y_i)$.  For the purposes of the present paper, it is more important to understand the statement of Lemma~\ref{lem:hs} than to remember the details of this construction.  Besides this lemma the only property of $\gamma(y_i)$ we will need is that its definition is independent of how elements of $C_i$ are labeled (Proposition \ref{prop:littlegammanatural}).

\begin{definition}
\label{def:littlegamma}
For any player $i$ in $\Gamma$ and any nonnegative $y_i\in\Rm^{C_i\times C_i}$, define the zero-sum game $\gamma(y_i)$ with strategy sets $C_M = C_m \eqdef C_i$ and utilities
\begin{equation*}
u_M(s_i,t_i) = -u_m(s_i,t_i) \eqdef \begin{cases}\sum_{r_i\neq t_i} y_i^{s_i,r_i}&\text{if }s_i = t_i, \\ -y_i^{s_i,t_i} & \text{otherwise}.\end{cases}
\end{equation*}
\end{definition}

\begin{lemma}[\cite{hs:ece}]
\label{lem:hs}
Fix a game $\Gamma$ and consider $\Gamma^0$.  If $y\in\Delta(C_m^0)$, then any strategy $\pi\in\maximin(\gamma(y_1))\times\cdots\times\maximin(\gamma(y_n))\subset\Delta(C_M^0)$ satisfies $u_M(\pi,y) = 0$.  In particular $v(\Gamma^0) = 0$, $\pi$ is good against $y$, and $\Delta^\Pi(\Gamma)$ is good.
\end{lemma}

\begin{proof}
Omitted.  See \cite{hs:ece} for a proof using the Minimax Theorem.
\end{proof}

\begin{theorem}[\cite{hs:ece}]
\label{thm:hs}
For any game $\Gamma$, the value $v(\Gamma^0)=0$, so $\maximin(\Gamma^0) = \ce(\Gamma)$ and a correlated equilibrium of $\Gamma$ exists.
\end{theorem}

\begin{proof}
Combining Lemma~\ref{lem:hs} and Proposition~\ref{prop:gamma0}, we get $\maximin(\Gamma^0) = \ce(\Gamma)$.  For existence, apply Proposition~\ref{prop:minimax} to $\Gamma^0$ with $\Sigma = \Delta^\Pi(\Gamma)$.
\end{proof}

This proof merits two remarks.  First of all, since $\clconv(\Delta^\Pi(\Gamma)) = \Delta(\Gamma)$, Proposition~\ref{prop:minimax} does not yield any benefit in this case over directly applying the Minimax Theorem to $\Gamma^0$.  Rather, we have used Proposition~\ref{prop:minimax} to illustrate our proof strategy for Theorems~\ref{thm:symexeqexist} and~\ref{thm:symordermexeqexist}, in which we use stronger versions of Lemma~\ref{lem:hs} to choose $\Sigma$ with $\clconv(\Sigma)\subsetneq\Delta(\Gamma)$.

Second, note that in this case we know that there is a maximin strategy of $\Gamma^0$ in the good set $\Delta^\Pi(\Gamma)$: this is just the statement of Nash's Theorem.  However, we cannot conclude this directly from the fact that $\Delta^\Pi(\Gamma)$ is a good set because of the remark after Proposition \ref{prop:minimax}.

\subsection{Groups acting on games}
\label{sec:sym}
In this section we recall the notion of a group acting on a game, as defined by Nash \cite{nash:ncg}.  All groups will be finite throughout.  In any group $e$ will denote the identity element.  The subgroup generated by group elements $g_1,\ldots, g_n$ will be denoted $\langle g_1,\ldots, g_n\rangle$.  For $n\in\Nm$ we will write $\Zm_n$ for the additive group of integers mod $n$ and $S_n$ for the symmetric group on $n$ letters.  We will use cycle notation to express permutations.  For example $\sigma = (1\ 2\ 3)(4\ 5)(6)$ is shorthand for
\begin{equation*}
\sigma(1) = 2,\ \sigma(2) = 3,\ \sigma(3)=1,\ \sigma(4) = 5,\ \sigma(5) = 4,\text{ and }\sigma(6)=6.
\end{equation*}

\begin{definition}
A \textbf{left action of the group $G$ on the set $X$} is a map $\cdot: G\times X\to X$ written with infix notation which satisfies the identity condition $e\cdot x = x$ and the associativity condition $g\cdot(h\cdot x) = (gh)\cdot x$.  A \textbf{right action of $G$ on $X$} is a map $\cdot: X\times G\to X$ such that $x\cdot e = x$ and $(x\cdot g)\cdot h = x\cdot (gh)$.  

We say that an action is \textbf{linear} if it extends to an action on an ambient vector space $V$ containing $X$ and the map $x\mapsto x\cdot g$ on $V$ is linear for all $g\in G$.  An $x\in X$ is \textbf{$G$-invariant} if $x\cdot g = x$ for all $g\in G$.  The set of $G$-invariant elements is denoted $X_G$.
\end{definition}

\begin{proposition}
\label{prop:average}
If $G$ acts linearly on the convex set $X$ then there is a map $\ave_G: X\to X_G$ given by $\ave_G(x) = \frac{1}{\abs{G}}\sum_{g\in G}x\cdot g$.  In particular if $X$ is nonempty then $X_G$ is nonempty.
\end{proposition}

\begin{proof}
For any $x\in X$, $\ave_G(x)$ is a convex combination of elements $x\cdot g\in X$, hence $\ave_G(x)\in X$.  For any $h\in G$ we have
\begin{equation*}
\begin{split}
\ave_G(x)\cdot h &= \left[\frac{1}{\abs{G}}\sum_{g\in G}x\cdot g\right]\cdot h = \frac{1}{\abs{G}}\sum_{g\in G}(x\cdot g)\cdot h = \frac{1}{\abs{G}}\sum_{g\in G}x\cdot (gh) \\ & = \frac{1}{\abs{G}}\sum_{g\in G}x\cdot g
= \ave_G(x),
\end{split}
\end{equation*}
where we have used linearity, the definition of a group action, and bijectivity of $g\mapsto gh$.
\end{proof}

A left action of $G$ on $X$ induces right actions on many function spaces defined on $X$.  For example $\Rm^X$ is the space of functions $X\to\Rm$.  For $y\in\Rm^X$ we can define $y\cdot g\in\Rm^X$ by $(y\cdot g)(x) = y(g\cdot x)$.  The condition that this is a right action of $G$ on $\Rm^X$ follows immediately from the fact that we began with a left action of $G$ on $X$.  For finite $X$ (the case of most interest to us), the same argument shows that $G$ acts on $\Delta(X)$ on the right.

\begin{definition}
We say that \textbf{a group $G$ acts on the game $\Gamma$} if the following conditions hold.  The group $G$ acts on the left on the set of players $I$ and $\bigsqcup_i C_i$, making $g\cdot s_i\in C_{g\cdot i}$ for $s_i\in C_i$.  Such actions automatically induce a left action of $G$ on $C = \prod_i C_i$ defined by $(g\cdot s)_{g\cdot i} = g\cdot s_i$.  We require that the utilities be invariant under the induced action on the right: $u_{g\cdot i} \cdot g = u_i$, i.e., $u_{g\cdot i}(g\cdot s) = u_i(s)$ for all $i\in I$, $s\in C$, and $g\in G$.  We say that $G$ is a \textbf{symmetry group of $\Gamma$} and call elements of $G$ \textbf{symmetries of $\Gamma$}.
\end{definition}

Note that an action of $G$ on a game can be fully specified by its action on $\bigsqcup_i C_i$ or on $C$.  One way to do this is to choose $G$ to be a subgroup of the symmetric group on $\bigsqcup_i C_i$ or $C$ satisfying the above properties.

\begin{definition}
\label{def:stabtrans}
The \textbf{stabilizer subgroup of player $i$} is $G_i \eqdef \{ g\in G \mid g\cdot i = i\}$, and acts on $C_i$ on the left.  We say that the action of $G$ is \textbf{player trivial} if $G_i = G$ for all $i$, or in other words if $g\cdot i = i$ for all $g$ and $i$.  We say that the action of $G$ is \textbf{player transitive} if for all $i,j\in I$ there exists $g\in G$ such that $g\cdot i = j$.
\end{definition}

We illustrate the notion of group actions on a game using four examples.

\savecounter{extrivial}
\begin{example}
\label{ex:trivial}
Let $\Gamma$ be any game and $G$ any group.  Define $g\cdot s = s$ for all $g\in G$ and $s\in C$.  This defines a player-trivial action of $G$ on $\Gamma$ called the trivial action.
\end{example}

\savecounter{exsymbimatrix}
\begin{example}
\label{ex:symbimatrix}
A two-player finite game is often called a \textbf{bimatrix game} because it can be described by two matrices $A$ and $B$, such that if player one plays strategy $i$ and player two plays strategy $j$ then their payoffs are $A_{ij}$ and $B_{ij}$, respectively.  If these matrices are square and $B = A^T$ then we call the game a \textbf{symmetric bimatrix game}.  One example is the game of chicken, which has $A = \left[\begin{smallmatrix}4 & 1 \\ 5 & 0\end{smallmatrix}\right] = B^T$.

To put this in the context of group actions defined above, let each player's strategy set be $C_1 = C_2 =  \{1,\ldots, m\}$ indexing the rows and columns of $A$ and $B$.  Define $g\cdot(i,j) = (j,i)$ for $(i,j)\in C$, so $g\cdot(g\cdot(i,j)) = (i,j)$.  The assumption $B = A^T$ is exactly the utility compatibility condition saying that this specifies an action of $G = \{e,g\}\cong\Zm_2$ on this game.  Of course, depending on the structure of $A$ and $B$ there may be other nontrivial symmetries as well.  The element $g$ swaps the players, so the action of $G$ is player transitive.
\end{example}

\begin{example}
Note that the condition that a bimatrix game be symmetric is \emph{not} that $A = A^T$ and $B=B^T$.  Indeed, such a game need not have any nontrivial symmetries.  For example, consider the game defined by $A = \left[\begin{smallmatrix}0 & 2 \\ 2 & 1\end{smallmatrix}\right]$ and $B=\left[\begin{smallmatrix}3 & 0 \\ 0 & 1\end{smallmatrix}\right]$.  The unique Nash equilibrium of this game is for player $1$ to play the mixed strategy $p = \left[\begin{smallmatrix}\frac{1}{4} & \frac{3}{4}\end{smallmatrix}\right]$ and player $2$ to play $q = \left[\begin{smallmatrix}\frac{1}{3} & \frac{2}{3}\end{smallmatrix}\right]$.  Since the equilibrium is unique, any symmetry of the game must induce a corresponding symmetry of the equilibrium by Nash's Theorem.  But the four entries of $p$ and $q$ are all distinct, so the only symmetry of this game is the trivial one.
\end{example}

\savecounter{exmp}
\begin{example}
\label{ex:mp}
Consider the game of matching pennies, whose utilities are shown in Table~\ref{tab:mp}.  The labels $H$ and $T$ stand for heads and tails, respectively, and the subscripts indicate the identities of the players for notational purposes.  This a bimatrix game, but it is not a symmetric bimatrix game in the sense of Example~\ref{ex:symbimatrix}.  %In fact even if we were allowed to relabel the strategies by permuting the rows and columns, we still could not make it a symmetric bimatrix game: a symmetric bimatrix game has outcomes (corresponding to the diagonal entries of $A$ and $B$) in which both players receive the same payoff.
\begin{table}
\begin{center}
\begin{tabular}{c|c|c|}
$(u_1,u_2)$ & $H_2$ & $T_2$ \\
\hline
$H_1$ & $(1,-1)$ & $(-1,1)$ \\
\hline
$T_1$ & $(-1,1)$ & $(1,-1)$ \\
\hline
\end{tabular}
\end{center}
\caption{Matching pennies.  Player $1$ chooses rows and player $2$ chooses columns.}
\label{tab:mp}
\end{table}

Nonetheless this game does have symmetries.  The easiest to see is the map $\sigma$ which interchanges the roles of heads and tails.  Letting $g$ be the permutation of $\bigsqcup_i C_i$ given in cycle notation as $g = (H_1\ T_1)(H_2\ T_2)$, we define $g\cdot s_i = g(s_i)$.  Another symmetry is the permutation $h = (H_1\ H_2\ T_1\ T_2)$. These satisfy $g^2 = e$ and $h^2 = g$, so $G = \langle h\rangle \cong\Zm_4$.  Note that $g$ acts on $I$ as the identity whereas $h$ swaps the players, so $G$ acts player transitively, whereas $\langle g\rangle\cong\Zm_2$ acts player trivially.
\end{example}

\savecounter{exnplayer}
\begin{example}
\label{ex:nplayer}
Now we consider an example of an $n$-player game with symmetries.  Throughout this example all arithmetic will be done mod $n$.  For simplicity in this example we will index the players using the members of $\Zm_n$ instead of the set $\{1,\ldots, n\}$.  Each player's strategy space will be $C_i = \Zm_n$ as well.  Define
\begin{equation*}
u_i(s_1,\ldots,s_n) = \begin{cases}1, & \text{when }s_i = s_{i-1}+1 \\ 0, & \text{otherwise}.\end{cases}
\end{equation*}

Then we can define a symmetry $g$ by $g(s_i) = s_i+1$, which increments each player's strategy by one mod $n$, but fixes the identities of the players.  Clearly $g$ is a permutation of order $n$.

We can define another symmetry $h$ which maps a strategy for player $i$ to the same numbered strategy for player $i+1$.  That is to say, $h$ acts on $C$ by cyclically permuting its arguments.  Again, $h$ is a permutation of order $n$.  Note that $g$ and $h$ commute, so together they generate a symmetry group $G \cong \Zm_n\times \Zm_n$.  Both $\langle h\rangle\cong\Zm_n$ and $G$ act player transitively, whereas $\langle g\rangle\cong\Zm_n$ acts player trivially.  If $n$ is composite and factors as $n=kl$ for $k,l>1$ then $\langle h^k\rangle\cong\Zm_l$ acts on $\Gamma$ but neither player transitively nor player trivially.
\end{example}

The left actions in the definition of a group action on a game induce linear right actions on function spaces such as $\Delta(\Gamma)\subsetneq\Rm^C$ and $\Delta^\Pi(\Gamma) \subsetneq \Rm^{\sqcup_i C_i}$.  The inclusion map $\Rm^{\sqcup_i C_i}\to\Rm^C$ is $G$-equivariant (commutes with the action of $G$), so with regard to this action it does not matter whether we choose to view $\Delta^\Pi(\Gamma)$ as a subset of $\Rm^{\sqcup_i C_i}$ or of $\Rm^C$.

Because of the utility compatibility conditions of a group action on a game, the actions on $\Delta(\Gamma)$ and $\Delta^\Pi(\Gamma)$ restrict to actions on the sets $\ce(\Gamma)$ and $\nash(\Gamma)$, respectively.  This allows us to define the $G$-invariant subsets $\Delta_G(\Gamma)$, $\Delta_G^\Pi(\Gamma)$, $\ce_G(\Gamma)$, and $\nash_G(\Gamma)$.  The action of the stabilizer subgroup $G_i$ on $C_i$ allows us to define the $G$-invariant subset $\Delta_{G_i}(C_i)$.

%\begin{definition}
%The set of \textbf{$G$-invariant} probability distributions on the joint action set of $\Gamma$ is
%\begin{equation*}
%\Delta_G(\Gamma) = \{ \pi\in\Delta(C)\mid \pi = \pi\cdot g\text{ for all }g\in G\}.
%\end{equation*}
%The set of \textbf{$G$-invariant independent probability distributions}, i.e., those in $\Delta_G(\Gamma)$ which factor as $\pi(s) = \pi_1(s_1)\cdots\pi_n(s_n)$, is denoted $\Delta_G^\Pi(\Gamma)$ (the $\Pi$ is for ``product'').
%\end{definition}

%We could just as well write $\Delta_G(C)$ instead of $\Delta_G(\Gamma)$ to denote elements of $\Delta(C)$ which are invariant under $G$, and at times we will.  

%\begin{definition}
%The set of \textbf{$G$-invariant correlated equilibria} is $\ce_G(\Gamma) = \ce(\Gamma) \cap \Delta_G(\Gamma)$.  The set of \textbf{$G$-invariant Nash equilibria} is $\nash_G(\Gamma) = \nash(\Gamma) \cap \Delta_G(\Gamma) = \ce(\Gamma)\cap \Delta_G^\Pi(\Gamma)$.
%\end{definition}

The main theorem we set out to prove is the following.  This theorem is most often applied in the case where $G$ is the trivial group, but Nash proved the general case in \cite{nash:ncg} and so shall we.

\begin{nashthm}
A game with symmetry group $G$ has a $G$-invariant Nash equilibrium.
\end{nashthm}

To prove this we will use Hart and Schmeidler's techniques in a new way.  We will show that certain classes of symmetric games have correlated equilibria with a much higher degree of symmetry than might be expected without knowledge of Nash's Theorem.  To illustrate what we mean, consider the following trivial improvement on Theorem~\ref{thm:hs}.

\begin{proposition}
\label{prop:symcorrexist}
A game with symmetry group $G$ has a $G$-invariant correlated equilibrium.
\end{proposition}

\begin{proof}
Apply Proposition~\ref{prop:average} to a correlated equilibrium, which exists by Theorem~\ref{thm:hs}.
%Let $\pi$ be any correlated equilibrium of the game, whose existence is guaranteed by, say, \cite{hs:ece}.  Then the definition of a symmetry of a game implies that $\pi\circ\hat{\sigma}$ is a correlated equilibrium for all $\sigma\in G$.  Define $\pi^* = \frac{1}{\lvert G\rvert}\sum_{\sigma\in G} \pi\circ \hat{\sigma}$.  Then for any $\tau\in G$ we have
%\begin{equation*}
%\pi^*\circ \hat{\tau} = \frac{1}{\lvert G\rvert}\sum_{\sigma\in G} \pi\circ (\hat{\sigma}\circ\hat{\tau}) =  \frac{1}{\lvert G\rvert}\sum_{\sigma\in G} \pi\circ  (\widehat{\sigma\circ\tau}) = 
%\frac{1}{\lvert G\rvert}\sum_{\sigma\in G} \pi\circ \hat{\sigma} = \pi^*,
%\end{equation*}
%since $\sigma\mapsto\sigma\circ\tau$ is a bijection from $G$ to itself.  Therefore $\pi^*$ is invariant under $\hat{\tau}$ for all $\tau\in G$.  The set of all correlated equilibria of a game is convex, so $\pi^*$ is a correlated equilibrium.
\end{proof}

A priori we might not expect correlated equilibria with a greater degree of symmetry than predicted by Proposition~\ref{prop:symcorrexist} to exist.  But viewing $G$-invariant Nash equilibria as correlated equilibria, we see that we can often guarantee much more.  Suppose we have an $n$-player game which has identical strategy sets for all players and which is symmetric under cyclic permutations of the players, such as the game in Example~\ref{ex:nplayer}.  Then Proposition~\ref{prop:symcorrexist} yields a correlated equilibrium $\pi$ which is invariant under cyclic permutations of the players, but need not be invariant under other permutations.  On the other hand the Nash equilibrium $\rho = (\rho_1,\ldots, \rho_n)$ given by Nash's Theorem satisfies $\rho_1 = \cdots = \rho_n$ so the corresponding product distribution $\pi(s_1,\ldots,s_n) = \rho_1(s_1)\cdots\rho_1(s_n)$ is a correlated equilibrium which is invariant under arbitrary permutations of the players.

\section{Exchangeable equilibria}
\label{sec:exeq}
In this section we prove the existence of correlated equilibria with this higher degree of symmetry, as well as a useful factorization property, without appealing to Nash's Theorem.

\subsection{Exchangeable distributions}
First we need the notion of an exchangeable probability distribution.  Our usage of the term ``exchangeable'' is slightly nonstandard but is closely related to the usual notion in the case when $G$ acts player transitively.

\begin{definition}
\label{def:deltax}
Viewing $\Delta_G^\Pi(\Gamma)$ as a nonconvex subset of the convex set $\Delta_G(\Gamma)$, we define the set of \textbf{$G$-exchangeable} probability distributions
\begin{equation*}
\Delta_G^X(\Gamma) \eqdef \conv\Delta_G^\Pi(\Gamma)\subseteq\Delta_G(\Gamma).
\end{equation*}
\end{definition}

We use the term ``exchangeable'' because of the important case where the $C_i$ are all equal and the group $G$ acts player transitively (e.g.\ in Example \ref{ex:nplayer}).  Then distributions in $\Delta_G^X(\Gamma)$ are invariant under arbitrary permutations of the players.  Furthermore, by De Finetti's Theorem\footnote{De Finetti's theorem states that the distribution of an infinite sequence of random variables $X_1, X_2,\ldots$ is invariant under permutations of finitely many of the random variables (\textbf{exchangeable}) if and only if it is a mixture of i.i.d.\ distributions \cite{s:tos}. } these are exactly the distributions which can be extended to exchangeable distributions on infinitely many copies of $C_1$, i.e., distributions invariant under permutations of finitely many indices.  De Finetti's Theorem will not play a direct role in our analysis; here it merely serves to motivate Definition~\ref{def:deltax}.

To get a feel for these sets, we will look at them in the context of some examples.

\usesavedcounter{extrivial}
\begin{example}[cont'd]
Since $G$ acts trivially we can ignore it entirely.  Not all distributions are independent so $\Delta_G^\Pi(\Gamma)\subsetneq\Delta_G(\Gamma) = \Delta(\Gamma)$, but $\Delta_G^X(\Gamma) = \Delta_G(\Gamma)$.  As we have seen, one inclusion is automatic.  To prove the reverse note that for any $s\in C$, $\delta_s = \delta_{s_1}\cdots\delta_{s_n}\in\Delta^\Pi(\Gamma) = \Delta_G^\Pi(\Gamma)$.  But for any $\pi\in\Delta(\Gamma)$ we can write $\pi = \sum_{s\in C} \pi(s)\delta_s$, and such a convex combination of the $\delta_s$ is in $\Delta_G^X(\Gamma)$ by definition.
\end{example}
\restorecounter

\usesavedcounter{exsymbimatrix}
\begin{example}[cont'd]
For a symmetric bimatrix game $\Gamma$ with $m$ strategies per player, we can view probability distributions over $C$ as $m\times m$ nonnegative matrices with entries summing to unity.  The nontrivial symmetry $g\in G$ acts by swapping the players.  From the definitions we see that $\Delta_G(\Gamma)$ consists of symmetric matrices and $\Delta_G^\Pi(\Gamma)$ of matrices which are outer products $xx^T$ for nonnegative column vectors $x\in\Rm^m$.  The elements of $\Delta_G^X(\Gamma) = \conv\Delta_G^\Pi(\Gamma)$ are exactly the (normalized) \textbf{completely positive matrices} studied in \cite{bsm:cpm}.  Clearly all such matrices are symmetric, elementwise nonnegative, and positive semidefinite; it turns out the converse holds if and only if $m\leq 4$ \cite{d:onnfrvsawnn}.
\end{example}
\restorecounter

\usesavedcounter{exmp}
\begin{example}[cont'd]
The map on $C$ induced by $h$ is the permutation
\begin{equation*}
((H_1,H_2)\ (T_1,H_2)\ (T_1,T_2)\ (H_1,T_2)).
\end{equation*}
In particular, a $G$-invariant probability distribution must assign equal probability to all four outcomes in $C_1\times C_2$.  There is only one such distribution and it is independent, so $\Delta_G^\Pi(\Gamma) = \Delta_G^X(\Gamma)=\Delta_G(\Gamma)$.
\end{example}
\restorecounter

\usesavedcounter{exnplayer}
\begin{example}[cont'd]
Recall that in this game there are $n$ players and the $C_i$ are the same for all $i$.  The group $G$ permutes the players cyclically.  Therefore the elements of $\Delta_G^\Pi(\Gamma)$ are invariant under arbitrary permutations of the players, hence so are the elements of $\Delta_G^X(\Gamma)$.  
(The converse statement is false; that is to say, there are probability distributions over $C$ which are invariant under arbitrary permutations of the players but are not in $\Delta_G^X(\Gamma)$.  This is analogous to the presence in Example~\ref{ex:symbimatrix} of symmetric elementwise nonnegative matrices which are not positive semidefinite, hence not completely positive.)  On the other hand, an element of $\Delta_G(\Gamma)$ need only be invariant under cyclic permutations of the players.  
\end{example}
\restorecounter

We close the section on exchangeable distributions with a characterization of $\Delta_G^X(\Gamma)$ which we will not need until Section \ref{sec:orderinfexeq} but which logically belongs here.  The characterization is a corollary of a more general convexity lemma which we state first.

\begin{lemma}
Let $f: K\to V$ be a continuous map from a compact set $K$ to a finite-dimensional real vector space $V$.  Extending $f$ by linearity yields a weakly continuous map $\Delta(K)\to V$ given by $\pi\mapsto \int f\,d\pi$ which we also call $f$.  The image of this map is $f(\Delta(K)) = \conv(f(K))$.
\end{lemma}

\begin{proof}
The extension $f$ is weakly continuous by definition.  Clearly $f(\Delta(K))$ is convex and contains $f(K)$, so one containment is immediate.  By linearity of integration, any linear inequality valid on $f(K)$ must be valid on $f(\Delta(K))$, so the reverse containment follows by a separating hyperplane argument (see Theorem~$3.1.1$ of \cite{karlin:tig} for details or Theorem $2.8$ of \cite{sop:slrcg} for an alternative topological argument).
\end{proof}

\begin{corollary}
\label{cor:xdistchar}
The linear extension of the inclusion map $\Delta_G^\Pi(\Gamma)\to \Delta_G(\Gamma)$ is weakly continuous and maps $\Delta(\Delta_G^\Pi(\Gamma))$ onto $\Delta_G^X(\Gamma)$.
\end{corollary}

\subsection{Exchangeable equilibria}
We are now ready to define exchangeable equilibria.  The proofs of the propositions in this section are direct algebraic manipulations and some are omitted.

\begin{definition}
\label{def:exeq}
The set of \textbf{$G$-exchangeable equilibria} of a game $\Gamma$ is
\begin{equation*}
\xe_G(\Gamma) \eqdef \ce(\Gamma)\cap\Delta_G^X(\Gamma).
\end{equation*}
When $G$ can be inferred from context we simply refer to \textbf{exchangeable equilibria}.
\end{definition}

It is immediate from the definitions that $\conv(\nash_G(\Gamma))\subseteq \xe_G(\Gamma)\subseteq \ce_G(\Gamma)$.  There are examples in which all of these inclusions are strict \cite{sop:xe1}, so proving non-emptiness of $\xe_G(\Gamma)$ does not immediately prove non-emptiness of $\nash_G(\Gamma)$.  Nonetheless, this is an important step and the main result of this section.

The proof that a $G$-exchangeable equilibrium exists proceeds along the same lines as the correlated equilibrium existence proof in Section~\ref{sec:hs}.  We again consider the zero-sum game $\Gamma^0$ and prove that a certain set is good in this game (Lemma~\ref{lem:exeqgood}).  The difference is that the action of $G$ yields a smaller good set, $\Delta_G^\Pi(\Gamma)$.  To prove this lemma we need the following three symmetry results, which have straightforward proofs.

\begin{proposition}
\label{prop:symgamma0}
If $G$ acts on $\Gamma$ then $G$ acts player trivially on $\Gamma^0$ by $g\cdot (s,(r_i,t_i)) = (g\cdot s,(g\cdot r_i,g\cdot t_i))$.
\end{proposition}

\begin{proposition}
\label{prop:symgood}
If $G$ acts player trivially on a zero-sum game, then a set $\Sigma\subseteq\Delta_G(C_M)$ is good if and only if it is good against $\Delta_G(C_m)$.
\end{proposition}

\begin{proof}
For all $g\in G$, $\sigma\in\Delta_G(C_M)$, and $\theta\in\Delta(C_m)$ we have $u_M(\sigma,\theta\cdot g) = u_M(\sigma\cdot g,\theta\cdot g) = u_M(\sigma,\theta)$, so $u_M(\sigma,\theta) = u_M(\sigma,\ave_G(\theta))$.
\end{proof}

\begin{proposition}
\label{prop:littlegammanatural}
The map $\maximin(\gamma(\cdot))$ is natural in the sense that if $\sigma: C_i\to C_j$ is a bijection and $y_i = y_j\circ(\sigma,\sigma)$, then composition with $\sigma$ maps $\maximin(\gamma(y_j))$ to $\maximin(\gamma(y_i))$.
\end{proposition}

%\begin{proposition}
%\label{prop:symlittlegamma}
%If $G$ acts on the game $\Gamma$ then the stabilizer subgroup $G_i$ acts on the joint action space $C_i\times C_i$ of the game in Definition~\ref{def:littlegamma} by acting on each factor separately.  If $y_i\in\Rm^{C_i\times C_i}_{G_i}$ then $G_i$ acts player trivially on $\gamma(y_i)$.
%\end{proposition}

%\begin{proposition}
%\label{prop:symmaximin}
%If $G$ acts player trivially on a zero-sum game, then the game has a $G$-invariant maximin strategy.
%\end{proposition}

%\begin{proof}
%Apply Proposition~\ref{prop:average} to the action of $G$ on the convex set $\maximin(\Gamma)$.
%%Let $\sigma\in\maximin(\Gamma)$.  For any $g\in G_M$ and any $\theta\in C_m$ we have $u_M(\sigma\cdot g,\theta\cdot g) = u_M(\sigma,\theta)\geq v(\Gamma)$.  But the map $\theta\mapsto \theta\cdot g$ is surjective, so $\sigma\cdot g\in\maximin(\Gamma)$.  The set $\maximin(\Gamma)$ is convex so $\ave_G(\sigma)\in\maximin_G(\Gamma)$.
%\end{proof}

%\begin{proposition}
%\label{prop:deltagpi}
%$\Delta_G^\Pi(\Gamma) = \{\pi_1\cdots\pi_n\mid \pi_i\in\Delta(C_i)\text{ and }\pi_i = \pi_j\cdot g\text{ whenever }g\cdot i = j\}$.
%\end{proposition}

%\begin{proof}These are the conditions the marginals of a $\pi\in\Delta_G(C)$ must satisfy.
%\end{proof}

\begin{lemma}
\label{lem:exeqgood}
If $G$ acts on the game $\Gamma$ then the set $\Delta_G^\Pi(\Gamma)$ is good in the zero-sum game $\Gamma^0$ of Definition \ref{def:gamma0}.
\end{lemma}

\begin{proof}
By Proposition~\ref{prop:symgamma0} and Proposition~\ref{prop:symgood}, it suffices to consider only $y\in\Delta_G(C_m^0)$, and show that there is a $\pi\in\Delta_G^\Pi(\Gamma)$ which is good against $y$.  Lemma~\ref{lem:hs} states that any $\pi\in S(y) \eqdef \maximin(\gamma(y_1))\times\cdots\times\maximin(\gamma(y_n))\subset\Delta^\Pi(\Gamma)$ is good against $y$.

By Proposition~\ref{prop:littlegammanatural} the action of $G$ on $\Delta^\Pi(\Gamma)$ restricts to a linear action of $G$ on $S(y)$ since $y\in\Delta_G(C_m^0)$.  Viewing $S(y)$ as a convex subset of $\Rm^{\sqcup_i C_i}$, Proposition~\ref{prop:average} shows the $G$-invariant subset $S_G(y)\subseteq\Delta_G^\Pi(\Gamma)$ is nonempty, so $\Delta_G^\Pi(\Gamma)$ is good.
\end{proof}

\begin{theorem}
\label{thm:symexeqexist}
A game with symmetry group $G$ has a $G$-exchangeable equilibrium.
\end{theorem}

\begin{proof}
By Theorem~\ref{thm:hs}, $\maximin(\Gamma^0) = \ce(\Gamma)$.  Lemma~\ref{lem:exeqgood} shows we can apply Proposition~\ref{prop:minimax} to $\Gamma^0$ with $\Sigma = \Delta_G^\Pi(\Gamma)$, proving that $\maximin(\Gamma^0)\cap\Delta_G^X(\Gamma) = \xe_G(\Gamma)$ is nonempty.
\end{proof}

It is worth contrasting the proof that symmetric correlated equilibria exist (Proposition~\ref{prop:symcorrexist}) with this proof.  Both involve averaging arguments to produce symmetric solutions.  The difference is that in the proof of Proposition~\ref{prop:symcorrexist} the averaging occurs within the set $\Delta(\Gamma)$, whereas in the case of Theorem~\ref{thm:symexeqexist} (in particular Lemma~\ref{lem:exeqgood}), the averaging occurs within $\Delta^\Pi(\Gamma)$, viewed as a convex subset of $\Rm^{\sqcup_i C_i}$.  By averaging within this smaller set, we guarantee that the resulting correlated equilibrium will have the additional symmetries discussed at the end of Section~\ref{sec:sym}.

The latter averaging argument requires a bit more care.  In particular, Proposition~\ref{prop:symcorrexist} is an immediate corollary of Theorem~\ref{thm:hs} on the existence of correlated equilibria.  On the other hand, to prove Theorem~\ref{thm:symexeqexist} we have to ``lift the hood'' on Theorem~\ref{thm:hs} and use Lemma~\ref{lem:hs} on good sets.  By doing so we exhibit a correlated equilibrium which we can prove lies in $\Delta_G^X(\Gamma)$ instead of just $\Delta_G(\Gamma)$.

\section{Higher order exchangeable equilibria}
\label{sec:higherexeq}
In this section we begin with a game $\Gamma$ and artificially add symmetries to produce two families of games $\mpow{\Gamma}$ and $\compow{\Gamma}$ with larger symmetry groups for each $m\in\Nm$.  Having constructed these games, we can exploit our knowledge of their structure to improve Theorem~\ref{thm:symexeqexist} and show that there are distributions which are simultaneously exchangeable equilibria of both $\mpow{\Gamma}$ and $\compow{\Gamma}$.  We call such distributions order $m$ exchangeable equilibria.

We then use a compactness argument to exhibit a distribution which is simultaneously an order $m$ exchangeable equilibrium for all $m\in\Nm$, called an order $\infty$ exchangeable equilibrium.  We will see in the next section that for player-transitive symmetry groups, an order $\infty$ exchangeable equilibrium is just a mixture of symmetric Nash equilibria.

Most of the work in this section consists of making the proper definitions.  Once that is done, the proofs are rather short.

\subsection{Powers of games}
\label{sec:powers}
To define order $m$ $G$-exchangeable equilibria we will need two notions of a power of a game $\Gamma$.  These are larger games in which multiple copies of $\Gamma$ are played simultaneously\footnote{
These definitions can be generalized to define two possible products of games, so that the powers we define reduce to $m$-fold products of a game with itself.  If we remove all mention of symmetry groups (so in particular exchangeable distributions and exchangeable equilibria are no longer defined) all the statements we make about Nash and correlated equilibria of these powers extend to corresponding statements about products with identical proofs.  We will not need this level of generality, however, so to avoid complicating notation we focus on powers.
}.  Throughout this section we will take as fixed a game $\Gamma$ with symmetry group $G$ and a number $m\in\Nm$.

\begin{definition}
For $m\in\Nm$, the \textbf{$m^{\text{th}}$ power of $\Gamma$}, denoted $\mpow{\Gamma}$, is a game in which $m$ independent copies of $\Gamma$ are played simultaneously.  More specifically, $\mpow{\Gamma}$ has $mn$ players labeled by pairs $i,j$, $1\leq i\leq n$, $1\leq j\leq m$, strategy spaces $\mpow{C}_{ij} \eqdef C_i$ for all $i,j$ with generic element $s_i^j$, and utilities $\mpow{u}_{ij}(s_1^1,\ldots, s_n^m) \eqdef u_i(s_1^j,s_2^j,\ldots, s_n^j)$.

The \textbf{contracted $m^{\text{th}}$ power of $\Gamma$}, denoted $\compow{\Gamma}$, is a game in which $m$ copies of $\Gamma$ are played simultaneously, but all by the same set of players.  Specifically, $\compow{\Gamma}$ has $n$ players, strategy spaces $\compow{C_i} \eqdef C_i^m$ with generic element $(s_i^1,\ldots,s_i^m)$ for all $i$, and utilities $\compow{u_i}(s_1^1,\ldots,s_n^m) \eqdef \sum_j u_i(s_1^j,s_2^j,\ldots, s_n^j)$.
\end{definition}

\begin{proposition}
\label{prop:powerdists}
Let $\Gamma$ be a game with symmetry group $G$ and fix $m\in\Nm$.  Then both powers $\mpow{\Gamma}$ and $\compow{\Gamma}$ are games with symmetry group $G\times S_m$ and they satisfy:
\begin{itemize}
\item $\Delta_{G\times S_m}\left(\mpow{\Gamma}\right) = \Delta_{G\times S_m}\left(\compow{\Gamma}\right)$
\item $\Delta_{G\times S_m}^\Pi\left(\mpow{\Gamma}\right) \subsetneq \Delta_{G\times S_m}^\Pi\left(\compow{\Gamma}\right)$
\item $\Delta_{G\times S_m}^X\left(\mpow{\Gamma}\right) \subseteq \Delta_{G\times S_m}^X\left(\compow{\Gamma}\right)$.
\end{itemize}
\end{proposition}

\begin{proof}
Both powers are invariant under arbitrary permutations of the copies and under symmetries in $G$ applied to all of the copies simultaneously.  In fact in the case of $\mpow{\Gamma}$ we can apply a different symmetry in $G$ to each copy independently so that $\mpow{\Gamma}$ is invariant under the larger group $G\wr S_m$ (the wreath product of $G$ and $S_m$), but we will not need this fact.

Since $G\times S_m$ acts on $\mpow{C}$ and $\compow{C}$ in the same way, we get the first equality.  The game $\mpow{\Gamma}$ has more players than $\compow{\Gamma}$, so $\Delta_{G\times S_m}^\Pi\left(\mpow{\Gamma}\right)$ has stronger independence conditions than $\Delta_{G\times S_m}^\Pi\left(\compow{\Gamma}\right)$, yielding the strict containment.  Taking convex hulls gives the final containment.
\end{proof}

Since $\Delta_{G\times S_m}(\mpow{\Gamma}) = \Delta_{G\times S_m}(\compow{\Gamma})$, we can compare the conditions for a distribution $\pi$ to be a correlated equilibrium of $\mpow{\Gamma}$ or $\compow{\Gamma}$.  We use the notation and terminology introduced for Proposition~\ref{prop:correqchar} to do so.

\begin{proposition}
\label{prop:powercorreqchar}
Let $(X_1^1,\ldots,X_n^m)$ be a random vector taking values in $C^m$ distributed according to $\pi\in\Delta_{G\times S_m}\left(\mpow{\Gamma}\right) = \Delta_{G\times S_m}\left(\compow{\Gamma}\right)$.  Then
\begin{itemize}
\item $\pi$ is a correlated equilibrium of $\mpow{\Gamma}$ if and only if $X_i^j$ is a best response (in $\Gamma$) to $\prob(X_{-i}^j\mid X_i^j)$ almost surely for all $i$ and $j$, and
\item $\pi$ is a correlated equilibrium of $\compow{\Gamma}$ if and only if $X_i^j$ is a best response (in $\Gamma$) to $\prob(X_{-i}^j\mid X_i^1,\ldots, X_i^m)$ almost surely for all $i$ and $j$.
\end{itemize}
\end{proposition}

\begin{proof}
By Proposition~\ref{prop:correqchar}, $\pi$ is a correlated equilibrium of $\mpow{\Gamma}$ if and only if $X_i^j$ is a best response in $\mpow{\Gamma}$ to $\prob(X^1,\ldots,X^{j-1},X_{-i}^j,X^{j+1},\ldots, X^m\mid X_i^j)$ almost surely for all $i$ and $j$.  But the utility of player $ij$ in $\mpow{\Gamma}$ is $u_i(X_1^j,\ldots,X_n^j)$, so player $ij$ can ignore $X_k^l$ for all $l\neq j$.

Now we consider when $\pi$ is a correlated equilibrium of $\compow{\Gamma}$.  Again by Proposition \ref{prop:correqchar}, this happens if and only if $(X_i^1,\ldots, X_i^m)$ is a best response in $\compow{\Gamma}$ to $\prob(X_{-i}^1,\ldots,X_{-i}^m\mid X_i^1,\ldots, X_i^m)$ almost surely for all $i$.  The utility of player $i$ in $\compow{\Gamma}$ is $\sum_j u_i(X_1^j,X_2^j,\ldots, X_n^j)$ and no $X_i^j$ appears in more than one term of this sum.  Thus the sum is maximized when each term is maximized independently.  That is to say $\pi\in\ce_{G\times S_m}(\compow{\Gamma})$ if and only if $X_i^j$ is a best response in $\Gamma$ to $\prob(X_{-i}^j\mid X_i^1,\ldots, X_i^m)$ almost surely for all $i$ and $j$.
\end{proof}

This characterization allows us to prove the following containments between equilibrium sets.  One can construct examples showing that in general none of the inclusions in this proposition can be reversed.  In particular, no containment holds between $\xe_{G\times S_m}\left(\mpow{\Gamma}\right)$ and $\xe_{G\times S_m}\left(\compow{\Gamma}\right)$ in either direction.  This is connected to the fact that the inclusion between the sets of correlated equilibria of $\mpow{\Gamma}$ and $\compow{\Gamma}$ goes in the opposite direction from the inclusion between the sets of Nash equilibria.

\begin{proposition}
\label{prop:powerprops}
The equilibria of $\mpow{\Gamma}$ and $\compow{\Gamma}$ satisfy
\begin{equation*}
\nash_{G\times S_m}\left(\mpow{\Gamma}\right) \subseteq \nash_{G\times S_m}\left(\compow{\Gamma}\right)\subseteq \ce_{G\times S_m}\left(\compow{\Gamma}\right)\subseteq\ce_{G\times S_m}\left(\mpow{\Gamma}\right).
\end{equation*}
\end{proposition}

\begin{proof}
We use Proposition \ref{prop:powercorreqchar}.  If $X_i^j$ are distributed according to $\pi\in\Delta_{G\times S_m}^\Pi(\mpow{\Gamma})$ then the $X_i^j$ are all independent, so the conditional distributions in Proposition \ref{prop:powercorreqchar} are equal to the corresponding unconditional distributions and both conditions are equivalent.  This proves the first containment.

The second containment follows because Nash equilibria are always correlated equilibria.  For the third containment, suppose $X_i^j$ is a best response to $\prob(X_{-i}^j\mid X_i^1,\ldots,X_i^m)$ almost surely.  Summing over possible values of $X_i^{-j}$ we get that $X_i^j$ is a best response to $\prob(X_{-i}^j\mid X_i^j)$ almost surely.
\end{proof}

For any $1\leq p<m$ we can define a projection map $\proj: \Delta_{G\times S_m}(\mpow{\Gamma})\to\Delta_{G\times S_p}(\mpowm{p}{\Gamma})$ which marginalizes out variables $X^{p+1},\ldots,X^m$.  This map respects the structure of all the sets of distributions mentioned in Proposition \ref{prop:powerdists}, i.e., it restricts to maps $\Delta_{G\times S_m}^\Pi(\mpow{\Gamma})\to\Delta_{G\times S_p}^\Pi(\mpowm{p}{\Gamma})$, $\Delta_{G\times S_m}^X(\compow{\Gamma})\to\Delta_{G\times S_p}^X(\compowm{p}{\Gamma})$, etc.  By Proposition \ref{prop:powercorreqchar} it also respects the equilibrium structure of these games in the sense that $\xe_{G\times S_m}(\mpow{\Gamma})$ maps into $\xe_{G\times S_p}(\mpowm{p}{\Gamma})$, $\nash_{G\times S_m}(\compow{\Gamma})$ maps into $\nash_{G\times S_p}(\compowm{p}{\Gamma})$, etc.

One can show that if $p=1$ then all of these maps are onto.  We use this fact only to motivate what follows and not in any of the arguments below, so we omit its proof.  In particular, this shows that $\proj(\xe_{G\times S_m}(\mpow{\Gamma})) = \xe_G(\Gamma) = \proj(\xe_{G\times S_m}(\compow{\Gamma}))$.  One can give an example in which $\conv(\nash_G(\Gamma))\subsetneq \xe_G(\Gamma)$ \cite{sop:xe1}, so neither of these projected sets of exchangeable equilibria approaches the convex hull of the Nash equilibria of $\Gamma$ as $m$ gets large.  We will see that taking the intersection $\xe_{G\times S_m}\left(\mpow{\Gamma}\right) \cap \xe_{G\times S_m}\left(\compow{\Gamma}\right)$ fixes this problem.

%\todo{Prove the claim about $p=1$?}

\subsection{Order $m$ $G$-exchangeable equilibria}
\label{sec:orderm}
By Propositions~\ref{prop:powerdists} and \ref{prop:powerprops} we have $\nash_{G\times S_m}(\mpow{\Gamma})\subseteq \xe_{G\times S_m}(\mpow{\Gamma})\cap \xe_{G\times S_m}(\compow{\Gamma})$.  Thus we expect the following definition not to be vacuous.

\begin{definition}
\label{def:ordermexeq}
The set of \textbf{order $m$ $G$-exchangeable equilibria of $\Gamma$} is
\begin{equation*}
\xe_G^m(\Gamma) \eqdef \xe_{G\times S_m}\left(\mpow{\Gamma}\right) \cap \xe_{G\times S_m}\left(\compow{\Gamma}\right),
\end{equation*}
or equivalently by Propositions~\ref{prop:powerdists} and \ref{prop:powerprops},
\begin{equation*}
\xe_G^m(\Gamma) \eqdef \Delta_{G\times S_m}^X\left(\mpow{\Gamma}\right)\cap \ce_{G\times S_m}\left(\compow{\Gamma}\right).
\end{equation*}
\end{definition}

\begin{figure}[htbp]
\begin{center}
\begin{tikzpicture}
[set/.style={},
incl/.style={right hook->,thick},
inclrev/.style={left hook->,thick}]%{draw=gray,inner sep=1pt,rounded corners,fill=gray!30,thick,inner sep=3pt}]
\node (nempow) at (0,0) [set] {$\nash_{G\times S_m}(\mpow{\Gamma})$};
\node (necompow) at (4,0) [set] {$\nash_{G\times S_m}(\compow{\Gamma})$};
\node (xem) at (2,2) [set] {$\xe_G^m(\Gamma)$};
\node (xempow) at (0,4) [set] {$\xe_{G\times S_m}(\mpow{\Gamma})$};
\node (xecompow) at (4,4) [set] {$\xe_{G\times S_m}(\compow{\Gamma})$};
\node (cempow) at (0,6) [set] {$\ce_{G\times S_m}(\mpow{\Gamma})$};
\node (cecompow) at (4,6) [set] {$\ce_{G\times S_m}(\compow{\Gamma})$};
\draw [incl] (nempow) to (necompow);
\draw [incl] (nempow) to (xem);
\draw [incl] (necompow) to (xecompow);
\draw [inclrev] (xem) to (xempow);
\draw [incl] (xem) to (xecompow);
\draw [inclrev] (xempow) to (cempow);
\draw [incl] (xecompow) to (cecompow);
\draw [inclrev] (cecompow) to (cempow);

\draw [->>,ultra thick] (6,3) to node [auto] {$\proj$} (8,3);

\node (ne) at (10,0) [set] {$\nash_G(\Gamma)$};
\node (xemproj) at (10,2) [set] {$\proj(\xe_G^m(\Gamma))$};
\node (xe) at (10,4) [set] {$\xe_G(\Gamma)$};
\node (ce) at (10,6) [set] {$\ce_G(\Gamma)$};
\draw [incl] (ne) to (xemproj);
\draw [incl] (xemproj) to (xe);
\draw [incl] (xe) to (ce);
\end{tikzpicture}
\caption{At the left is a summary of the containments between equilibrium sets of the powers $\mpow{\Gamma}$ and $\compow{\Gamma}$ proven in Section \ref{sec:powers}.  An arrow $A\hookrightarrow B$ indicates $A\subseteq B$.  Under the marginalization map $\proj: \Delta_{G\times S_m}(\mpow{\Gamma})\to\Delta_G(\Gamma)$ each of these sets maps onto the set at the same height on the right.}
\label{fig:eqincl}
\end{center}
\end{figure}
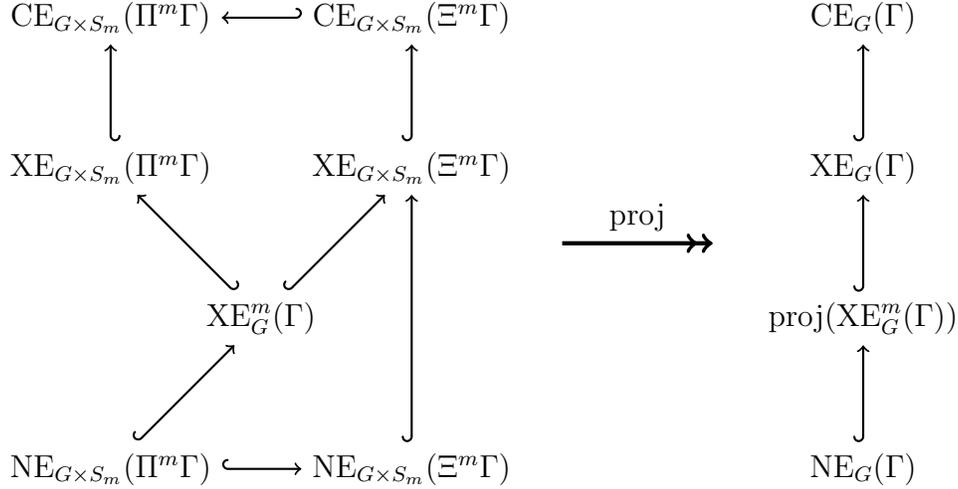

The relationship between $\xe_G^m(\Gamma)$ and the sets of equilibria of the $m^{\text{th}}$ powers is summarized in Figure \ref{fig:eqincl}.  We now prove order $m$ $G$-exchangeable equilibria exist.

\begin{lemma}
\label{lem:ordermexeqgood}
If $G$ acts on the game $\Gamma$ then the set $\Delta_{G\times S_m}^\Pi(\mpow{\Gamma})$ is good in the zero-sum game $(\compow{\Gamma})^0$ of Definition \ref{def:gamma0}.
\end{lemma}

\begin{proof}
By Lemma~\ref{lem:exeqgood}, $\Sigma \eqdef \Delta_{G\times S_m}^\Pi(\compow{\Gamma})$ is good.  The utilities in $\compow{\Gamma}$ are additively separable, so any mixed strategy $\sigma\in\Sigma$ is payoff equivalent for the maximizer in $(\compow{\Gamma})^0$ to a mixed strategy $\sigma'\in\Sigma' \eqdef\Delta_{G\times S_m}^\Pi(\mpow{\Gamma})$ given by the product of the marginals of $\sigma$.  We can apply Proposition \ref{prop:stratequivgood} to $\Sigma$ and $\Sigma'$.
\end{proof}

\begin{theorem}
\label{thm:symordermexeqexist}
A game with symmetry group $G$ has an order $m$ $G$-exchangeable equilibrium for all $m\in\Nm$.
\end{theorem}

\begin{proof}
By Theorem~\ref{thm:hs}, $\maximin((\compow{\Gamma})^0) = \ce(\compow{\Gamma})$.  Lemma~\ref{lem:ordermexeqgood} shows we can apply Proposition~\ref{prop:minimax} to $(\compow{\Gamma})^0$ with $\Sigma = \Delta_{G\times S_m}^\Pi(\mpow{\Gamma})$, so $\maximin((\compow{\Gamma})^0)\cap\Delta_{G\times S_m}^X(\mpow{\Gamma}) = \xe_G^m(\Gamma)$ is nonempty.
\end{proof}

For $1\leq p<m$, the marginalization map sends $\xe_{G\times S_m}(\mpow{\Gamma})$ into $\xe_{G\times S_p}(\mpowm{p}{\Gamma})$ and $\xe_{G\times S_m}(\compow{\Gamma})$ into $\xe_{G\times S_p}(\compowm{p}{\Gamma})$.  Therefore it sends $\xe_G^m(\Gamma)$ into $\xe_G^p(\Gamma)$.  Projecting the order $m$ exchangeable equilibria into $\Delta_G(\Gamma)$ for all $m\in\Nm$ we obtain
\begin{equation*}
\nash_G(\Gamma)\subseteq\conv(\nash_G(\Gamma))\subseteq\cdots\subseteq \proj(\xe_G^3(\Gamma))\subseteq \proj(\xe_G^2(\Gamma))\subseteq \xe_G(\Gamma)\subseteq\ce_G(\Gamma).
\end{equation*}
This raises two natural questions:  can we prove directly that $\bigcap_{m=1}^\infty \proj(\xe_G^m(\Gamma))$ is non\-empty? and does this intersection equal $\conv(\nash_G(\Gamma))$?  We will take up these two questions, respectively, in the following two sections.

\subsection{Order $\infty$ $G$-exchangeable equilibria}
\label{sec:orderinfexeq}
Next we use a compactness argument to prove existence of an order $\infty$ $G$-exchangeable equilibrium, a distribution which is in some sense an order $m$ $G$-exchangeable equilibrium for all finite $m$.  As we have defined them the $\xe_G^m(\Gamma)$ are distributions over different numbers of copies of $C$, so they are not directly comparable and we can't just take their intersection.  We could project them all into $\Delta_G(\Gamma)$ and take the intersection there as mentioned above, but this would destroy some structure.  Analytically it will be more convenient to view these sets within a larger space.

To take the intersection properly, we will transport all the $\xe_G^m(\Gamma)$ into $\Delta(\Delta_G^\Pi(\Gamma))$.  Since $S_m$ acts transitively on the copies of the game in $\mpow{\Gamma}$, an element $\rho\in\Delta_{G\times S_m}^\Pi(\mpow{\Gamma})$ satisfies $\rho_i^j = \rho_i^k$ for all $i$,$j$, and $k$.  Therefore the diagonal map $\Delta_G^\Pi(\Gamma)\to\Delta_{G\times S_m}^\Pi(\mpow{\Gamma})$ which sends a distribution to the product of $m$ independent copies of itself is a homeomorphism.  This extends to a homeomorphism $\Delta(\Delta_G^\Pi(\Gamma))\to\Delta(\Delta_{G\times S_m}^\Pi(\mpow{\Gamma}))$ of the corresponding spaces of distributions.  If we compose this with the surjective map $\Delta(\Delta_{G\times S_m}^\Pi(\mpow{\Gamma}))\to\Delta_{G\times S_m}^X(\mpow{\Gamma})$ given by Corollary~\ref{cor:xdistchar}, we get a surjective map $\mu_m: \Delta(\Delta_G(\Gamma))\to \Delta_{G\times S_m}^X(\mpow{\Gamma})$.

Define the inverse image sets $A_m =  \mu_m^{-1}(\xe_G^m(\Gamma))$.  Elements of $A_m$ are representations of order $m$ $G$-exchangeable equilibria as mixtures of independent $G$-invariant distributions.

\begin{definition}
The set of order $\infty$ $G$-exchangeable equilibria is $\xe_G^\infty(\Gamma) \eqdef \bigcap_{m=1}^\infty A_m$.
\end{definition}

\begin{theorem}
\label{thm:symorderinfexeqexist}
A game with symmetry group $G$ has an order $\infty$ $G$-exchangeable equilibrium.
\end{theorem}

\begin{proof}
%Endow $\Delta(\Delta_G^\Pi(\Gamma))$ with the topology of weak convergence, which makes it into a compact metric space since $\Delta_G^\Pi(\Gamma)$ is ($11.5.4$ and $11.5.5$ in \cite{d:rap}).  For any $(s^1,\ldots,s^m)\in C^m$ the map $\Delta_G^\Pi(\Gamma)\to\Rm$ given by $R\mapsto R(s^1)\cdots R(s^m)$ is a polynomial, hence continuous, so the map $\mu_m$ is continuous by definition of weak convergence.
Each set $\xe_G^m(\Gamma)$ is convex and closed by definition and nonempty by Theorem~\ref{thm:symordermexeqexist}.  The map $\mu_m$ is linear, weakly continuous, has a compact Hausdorff domain, and is surjective.  Therefore each $A_m$ is convex, compact, and nonempty.

For $1\leq p<m$, the composition $\mu_p\circ\mu_m^{-1}$ is a well-defined map $\Delta_{G\times S_m}^X(\mpow{\Gamma})\to\Delta_{G\times S_p}^X(\mpowm{p}{\Gamma})$ which coincides with the marginalization map discussed above.  This map sends $\xe_G^m(\Gamma)$ into $\xe_G^p(\Gamma)$ as per the discussion at the end of Section~\ref{sec:orderm}, so the $A_m$ are nested $A_1\supseteq A_2 \supseteq A_3\supseteq \ldots$.  Thus they have nonempty, convex, compact intersection $\xe_G^\infty(\Gamma)$.
\end{proof}

\section{Nash's Theorem}
\label{sec:nash}
\subsection{The player-transitive case}
\label{sec:nashtrans}
\begin{theorem}
\label{thm:orderinfexeqchar}
If $G$ acts player transitively on $\Gamma$, then $\xe_G^\infty(\Gamma) = \Delta(\nash_G(\Gamma))$.
\end{theorem}

\begin{proof}
If $\sigma\in\nash_G(\Gamma)$ then $\mu_m(\delta_\sigma)\in\nash_{G\times S_m}(\mpow{\Gamma})\subseteq\xe_G^m(\Gamma)$, so $\delta_\sigma\in A_m$ for all $m$ and $\delta_\sigma\in \xe_G^\infty(\Gamma)$.  But $\xe_G^\infty(\Gamma)$ is convex and weakly closed, so $\Delta(\nash_G(\Gamma)) = \clconv\{\delta_\sigma\mid \sigma\in\nash_G(\Gamma)\}\subseteq \xe_G^\infty(\Gamma)$.

For the converse let $\mathcal{R}$ be a random variable taking values in $\Delta_G^\Pi(\Gamma)$ distributed according to $\pi\in \xe_G^\infty(\Gamma)$.  Let $X_i^j$, $1\leq i\leq n$, $1\leq j<\infty$, be random variables taking values in $C_i$ with distribution $\mathcal{R}_i$ which are conditionally independent given $\mathcal{R}$.  We must show that $\mathcal{R}_i$ is a best response to $\mathcal{R}_{-i}$ almost surely.  We will do this by approximating $\mathcal{R}_i$ and $\mathcal{R}_{-i}$ in terms of the $X_i^j$.

For each $k\in\Nm$ the finite collection of random variables $X_i^j$ with $j\leq k$ is distributed according to $\mu_k(\pi)$ by construction; here we implicitly use the fact that $\mu_k(\pi)$ is an order $\infty$ $G$-exchangeable equilibrium, so $\mu_k(\pi)\in\Delta_{G\times S_m}^X(\mpow{\Gamma})$.  Furthermore we have $\mu_k(\pi)\in\ce_{G\times S_k}(\compowm{k}{\Gamma})$, so  Proposition~\ref{prop:powercorreqchar} states that for any $1\leq j\leq k$ the strategy $X_i^j$ is a best response to the random conditional distribution $\prob(X_{-i}^j \mid X_i^1,\ldots, X_i^k)$ almost surely.

Since $\mu_k(\pi)$ is symmetric, $\prob(X_{-i}^j \mid X_i^1,\ldots, X_i^k) \equiv \prob(X_{-i}^1 \mid X_i^1,\ldots, X_i^k)$ for all $i$, $j$, and $k$.  We define $\mathcal{P}_i^k$ to be this common random conditional distribution.  Let $\mathcal{Y}_i^j$ be the random variable taking values in $\Delta(C_i)$ which is the empirical distribution of $X_i^1,\ldots, X_i^j$.  Then $\mathcal{Y}_i^j$ is a best response to $\mathcal{P}_i^k$ whenever $j\leq k$.  We will show that $\mathcal{Y}_i^j$ and $\mathcal{P}_i^k$ converge to $\mathcal{R}_i$ and $\mathcal{R}_{-i}$, respectively, as $j$ and $k$ go to infinity.

Let $\Sigma_i$ be the completion of the $\sigma$-algebra generated by $X_i^1,X_i^2,\ldots$ and define $\mathcal{P}_i^\infty \eqdef \prob(X_{-i}^1\mid \Sigma_i)$.  Then $\mathcal{P}_i^k\rightarrow \mathcal{P}_i^\infty$ almost surely as $k$ goes to infinity (Theorem $10.5.1$ in \cite{d:rap}).  Therefore $\mathcal{Y}_i^j$ is a best response to $\mathcal{P}_i^\infty$ for all $j$.  By the strong law of large numbers, $\mathcal{Y}_i^j$ converges almost surely to $\mathcal{R}_i$ as $j$ goes to infinity, so $\mathcal{R}_i$ is a best response to $\mathcal{P}_i^\infty$.  Furthermore, $\mathcal{R}_i$ is measurable with respect to $\Sigma_i$ because the $\mathcal{Y}_i^j$ are.

The $X_i^j$ are conditionally independent given $\mathcal{R}$, so we have $\mathcal{P}_i^\infty = \expect(\prob(X_{-i}^1\mid \mathcal{R})  \mid \Sigma_i)$.  Since $G$ acts player transitively, for any player $j$ we have $\mathcal{R}_j = \mathcal{R}_i\cdot g$ for some $g\in G$, hence $\mathcal{R}_j$ is measurable with respect to $\Sigma_i$ and so is $\mathcal{R}$.  In particular $\prob(X_{-i}^1\mid \mathcal{R})$ is measurable with respect to $\Sigma_i$ and we obtain
\begin{equation*}
\mathcal{P}_i^\infty = \expect(\prob(X_{-i}^1\mid \mathcal{R})  \mid \Sigma_i)=\prob(X_{-i}^1\mid \mathcal{R}) = \mathcal{R}_{-i}.
\end{equation*}
This shows that $\mathcal{R}_i$ is a best response to $\mathcal{R}_{-i}$ almost surely for all $i$, so $\mathcal{R}\in\nash_G(\Gamma)$ almost surely and $\pi\in\Delta(\nash_G(\Gamma))$.
\end{proof}

If $G$ is the trivial group one can show that $\mu_1(\xe_G^\infty(\Gamma)) = \ce(\Gamma)$ and $\mu_1(\Delta(\nash_G(\Gamma))) = \conv(\nash(\Gamma))$.  These sets are different for some games (e.g., chicken), so the above theorem can fail without the player-transitivity assumption.

\begin{nashthm}[player-transitive case]
A game with player-transitive symmetry group $G$ has a $G$-invariant Nash equilibrium.
\end{nashthm}

\begin{proof}Combine Theorems~\ref{thm:symorderinfexeqexist} and~\ref{thm:orderinfexeqchar}, noting that $\Delta(\emptyset)=\emptyset$.
\end{proof}

\subsection{Arbitrary symmetry groups}
\label{sec:finishproof}
In this section we show how to embed an arbitrary game $\Gamma$ with symmetry group $G$ in a game $\Gamma^\Sym$ with a player-transitive symmetry group, preserving the existence of $G$-invariant Nash equilibria.  This allows us to drop the player-transitivity assumption from the previous section, proving Nash's Theorem in full generality.

There are a variety of ways to symmetrize games.  The one we have chosen is a natural $n$-player generalization of von Neumann's tensor-sum symmetrization discussed in \cite{gkt:osg}.  The idea is that each of the $n$ players in $\Gamma^\Sym$ plays all the roles of the players in $\Gamma$ simultaneously.  The players in $\Gamma^\Sym$ play $n!$ copies of $\Gamma$, one for each assignment of players in $\Gamma^\Sym$ to roles in $\Gamma$.  A player's utility in $\Gamma^\Sym$ is the sum of his utilities over the copies.

\begin{definition}
Given an $n$-player game $\Gamma$ with strategy sets $C_i$ and utilities $u_i$ we define its \textbf{symmetrization} $\Gamma^\Sym$ to be the $n$-player game with strategy sets $C_i^\Sym \eqdef C$ (with typical strategy $s^i = (s^i_1,\ldots,s^i_n)$) and utilities
\begin{equation*}
u_i^\Sym(s) \eqdef \sum_{\tau\in S_n} u_{\tau(i)} \left(d(\tau\star s)\right),
\end{equation*}
where $s = (s^1,\ldots, s^n) \in C^\Sym = C^n$, $\star: S_n\times C^\Sym\to C^\Sym$ is defined by $(\tau\star s)^k \eqdef s^{\tau^{-1}(k)}$, and $d: C^\Sym\to C$ is defined by $[d(s)]_k \eqdef s_k^k$.
%\begin{equation*}
%u_i^\Sym(s^1,\ldots, s^n) = \sum_{\tau\in S_n} u_{\tau(i)}\left(s_1^{\tau^{-1}(1)},\ldots,s_n^{\tau^{-1}(n)}\right),
%\end{equation*}
%where $s^i \in C_i^\Sym = C$.
\end{definition}

We now show that $\Gamma^\Sym$ is a game with player-transitive symmetry group.  We will use $\star$ to denote the action on $\Gamma^\Sym$ to distinguish it from the action $\cdot$ on $\Gamma$.

\begin{proposition}
\label{prop:symmetrization}
If $\Gamma$ is a game with symmetry group $G$ then $\Gamma^\Sym$ is a game with player-transitive symmetry group $G\times S_n$, where $\sigma\in S_n$ acts by $\star$ as defined above and $g\in G$ acts by
\begin{equation*}
g\star(s^1,\ldots,s^n)\mapsto (g\cdot s^1,\ldots, g\cdot s^n).
\end{equation*}
\end{proposition}

\begin{proof}
Note that $\star$ defines an action of $G$ on $C^\Sym$.  Also, for $\sigma,\tau\in S_n$ we have
\begin{equation*}
(\tau\star(\sigma\star s))^k = (\sigma\star s)^{\tau^{-1}(k)} = s^{\sigma^{-1}(\tau^{-1}(k))} = s^{(\tau\sigma)^{-1}(k)} = ((\tau\sigma)\star s)^k,
\end{equation*}
so $\star$ is an action of $S_n$ on $C^\Sym$ as well.  These actions commute, so together they define an action $\star$ of $G\times S_n$ on $C^\Sym$.  Note that the induced actions on players are $g\star i = i$ and $\sigma\star i = \sigma(i)$.

To show that this is an action of $G\times S_n$ on $\Gamma^\Sym$ it suffices to show that the utilities of $\Gamma^\Sym$ are invariant under the action of any $\sigma\in S_n$ and any $g\in G$.  To see the former, let $\sigma\in S_n$.  Then we have
\begin{equation*}
\begin{split}
u_{\sigma\star i}^\Sym(\sigma\star s) & = \sum_{\tau\in S_n} u_{\tau(\sigma(i))}\left(d(\tau\star(\sigma\star s))\right) = \sum_{\tau\in S_n} u_{(\tau\sigma)(i)}\left(d((\tau\sigma)\star s\right) = \sum_{\tau\in S_n} u_{\tau(i)} \left(d(\tau\star s)\right) \\ & = u_i^\Sym(s),
\end{split}
\end{equation*}
where we have used in the penultimate equation the fact that $S_n$ is a group, so the map $\tau\mapsto\tau\sigma$ is a bijection.  To see the latter, let $g\in G$ and let $\gamma\in S_n$ be the permutation induced by $g$ on the set of players in $\Gamma$.  Then we have $d(g\star s) = g\cdot d(\gamma^{-1}\star s)$, so
\begin{equation*}
\begin{split}
u_{g\star i}^\Sym(g\star s) & = \sum_{\tau\in S_n} u_{\tau(i)} \left(d(\tau\star (g\star s))\right)  = \sum_{\tau\in S_n} u_{\tau(i)} \left(d(g\star(\tau\star s))\right) \\
& = \sum_{\tau\in S_n} u_{\tau(i)} \left(g\cdot d(\gamma^{-1}\star(\tau\star s))\right) = \sum_{\tau\in S_n} u_{(\gamma^{-1}\tau)(i)} \left(d((\gamma^{-1}\tau)\star s)\right)\\
&  = \sum_{\tau\in S_n} u_{\tau(i)} \left(d(\tau\star s)\right) = u_i^\Sym(s),
\end{split}
\end{equation*}
where the fourth equation follows because $g$ is a symmetry of $\Gamma$.  Clearly $S_n$ acts transitively on the set of players.
\end{proof}

\begin{nashthm}
A game with symmetry group $G$ has a $G$-invariant Nash equilibrium.
\end{nashthm}

\begin{proof}
Let $\Gamma$ be a game with symmetry group $G$.  Then $\Gamma^\Sym$ is a game with player-transitive symmetry group $G\times S_n$ by Proposition~\ref{prop:symmetrization}, so it has a $(G\times S_n)$-symmetric Nash equilibrium by the player-transitive version of Nash's Theorem.  By definition of the action of $G\times S_n$ on $\Gamma^\Sym$, this Nash equilibrium is of the form $(\rho,\ldots, \rho)$, with $\rho\in\Delta_G(\Gamma)$.  Notice that for each player $i$, each utility $u_k^\Sym(s^1,\ldots,s^n)$ is a sum of functions which only depend on $s^i_j$ for a single value of $j$.  Thus $\rho$ is payoff equivalent to the product of its marginals $\rho_1\times\cdots\times\rho_n\in\Delta_G^\Pi(\Gamma)$.  Therefore we can take the Nash equilibrium $(\rho,\ldots,\rho)$ to be such that $\rho\in\Delta_G^\Pi(\Gamma)$ by Proposition~\ref{prop:stratequivnash}.

It remains to verify that $\rho\in\nash_G(\Gamma)$.  For any $s^i\in C$ we can compute
\begin{equation*}
\begin{split}
u_i^\Sym(\rho,\ldots,\rho,s^i,\rho,\ldots,\rho) & = \sum_{\tau\in S_n} u_{\tau(i)} \left(\rho_1,\ldots, \rho_{\tau(i)-1},s^i_{\tau(i)},\rho_{\tau(i)+1},\ldots,\rho_n\right) \\
& = (n-1)!\sum_{j=1}^n u_j(\rho_1,\ldots,\rho_{j-1},s^i_j,\rho_{j+1},\ldots,\rho_n).
\end{split}
\end{equation*}
For each value of $j$ we can vary the $s^i_j$ component of $s^i$ independently and it is a best response for player $i$ to play $\rho$ in $\Gamma^\Sym$ if the rest of the players play $\rho$, so we must have
\begin{equation*}
u_j(\rho_1,\ldots,\rho_{j-1},s_j,\rho_{j+1},\ldots,\rho_n)\leq u_j(\rho)
\end{equation*}
for all players $j$ and all $s_j\in C_j$, i.e., $\rho\in\nash_G(\Gamma)$.
\end{proof}

\section{Conclusion}
\label{sec:conclusions}
We have shown that by studying group actions on games and introducing the notion of exchangeable equilibrium, we can prove Nash's Theorem.  To the authors' knowledge, this is the first proof of this theorem which uses convexity-based methods (i.e., the minimax theorem).  Previous proofs use path-following arguments or fixed-point theorems, which are essentially equivalent to path-following arguments by Sperner's Lemma.

This new proof invites new approaches for computing or approximating Nash equilibria.  One can rewrite the existence proof above for (order $m$) exchangeable equilibria in terms of linear programs and separation arguments instead of the Minimax Theorem and apply the ellipsoid algorithm, just as Papadimitriou has done for Hart and Schmeidler's proof of the existence of correlated equilibria \cite{p:ccempg}.  This shows that exchangeable equilibria can be computed in polynomial time, at least under some assumptions on the parameters.  For example, order $m$ exchangeable equilibria of symmetric bimatrix games can be computed in polynomial time for fixed $m$.

We have seen that in the player-transitive case order $m$ exchangeable equilibria converge to convex combinations of Nash equilibria as $m$ goes to infinity.  There are a variety of ways one could imagine ``rounding'' exchangeable equilibria to try to produce  approximate Nash equilibria.  We leave the analysis of such procedures, along with the question of which assumptions on $G$ allow computation of exchangeable equilibria in polynomial time, for future work.

The power of these methods suggests that exchangeable equilibria should not merely be viewed as a step on the way to Nash equilibria.  Rather, they deserve further study in their own right.  We consider several interpretations of exchangeable equilibria and the applications they suggest in \cite{sop:xe1}.

\bibliographystyle{plain}
\bibliography{../../references}
\end{document}

%\markright{LIDS Technical Report $2760$}

\maketitle

\thispagestyle{headings}

\pagestyle{plain}

The goal of this note is to explain a problem with the proof of Nash's theorem in \cite{spo:nashsthm}, which is the main document attached below.  We will use the notation of that paper throughout without further comment, with the exception that each document has its own section numbers and list of references.  We are indebted to Sergiu Hart and Eran Shmaya for their careful study of \cite{spo:nashsthm} which led to their simultaneous discovery of this error.  So far the error has not been fixed, but many of the results and techniques of the paper remain valid, so we will continue to make it available online.

\section{The problem}
\label{sec:intro}
The error is in our proof of Lemma $4.6$, which is the main driver behind our assertion that order $m$ exchangeable equilibria exist (Theorem $4.7$).  Of course they do exist; this is the main benefit of setting out to prove something you already know to be true.  To our knowledge everything before and after this step remains unaffected, but this is an important step in the middle of the proof and logically it must be patched if we are to reach the desired conclusion.

The particular problem is our implicit use of the ``fact'' below.  In the incorrect proof of Lemma $4.6$ of \cite{spo:nashsthm}, this ``fact'' was applied with $\compow{\Gamma}$ in place of $\Gamma$ and appropriate choices of $\sigma\in\Delta_{G\times S_m}^\Pi(\compow{\Gamma})$ and $\tau\in\Delta_{G\times S_m}^\Pi(\mpow{\Gamma})$.

\begin{fact}
Let $\sigma, \tau\in\Delta^\Pi(\Gamma)$ be two strategy profiles such that $\sigma_i$ is payoff equivalent to $\tau_i$ in $\Gamma$ for all $i$.  Then $\sigma$ and $\tau$ are payoff equivalent when viewed as strategies for the maximizer in the associated zero-sum game $\Gamma^0$.
\end{fact}

This may sound reasonable at first because it seems that the difference between payoff-equivalent strategies should not be game-theoretically significant.  In fact it is false!  And not just for odd counterexamples: it is false generically.

\begin{example}
\label{ex:correqjump}
To see how this can fail, let $\Gamma$ be the following symmetric bimatrix game:
\begin{equation*}
A = B^T = \begin{bmatrix}1 & 1 & 1 \\ 1 & 1 & 1 \\ 0 & 0 & 2 \end{bmatrix}.
\end{equation*}
Let player $i$'s strategy set be $C_i = \{a_i, b_i, c_i\}$.  The first two columns of $A$ are equal, as are the first two rows, so $a_1$ is payoff equivalent to $b_1$.  Symmetrically, $a_2$ is payoff equivalent to $b_2$.  In the associated zero-sum game $\Gamma^0$ we have
\begin{equation*}
u_M^0((a_1,a_2),(a_1,c_1)) = 1 \neq 0 = u_M^0((b_1,b_2),(a_1,c_1)),
\end{equation*}
so $(a_1,a_2)$ and $(b_1,b_2)$ are not payoff equivalent in $\Gamma^0$.

Why does this happen?  The reason is that when we consider correlated strategies, we are interested not only in what utility our recommendations give us, but also in what information they give.  While we may not care from a payoff perspective whether strategy $a_i$ or $b_i$ has been suggested, the two possibilities may allow us to make completely different inferences about the behavior of our opponents.  It is this information which may be payoff-relevant.

To see this in action, consider the following two distributions over outcomes of $\Gamma$:
\begin{equation*}
W^1 = \frac{1}{4}\begin{bmatrix}1 & 0 & 1 \\ 0 & 0 & 0 \\ 1 & 0 & 1\end{bmatrix}\qquad\text{and}\qquad W^2 = \frac{1}{4}\begin{bmatrix}1 & 0 & 0 \\ 0 & 0 & 1 \\ 0 & 1 & 1\end{bmatrix}.
\end{equation*}
One can verify that $W^1$ is a correlated equilibrium, and $W^2$ is obtained from $W^1$ by moving some mass from $(a_1,c_2)$ and $(c_1,a_2)$ to $(b_1,c_2)$ and $(c_1,b_2)$, which are profiles of payoff equivalent strategies.  Nonetheless, $W^2$ is not a correlated equilibrium.  If player $1$ receives the recommendation $b_1$ then he knows with probability one that player $2$ has received the recommendation $c_2$, hence it is in the interest of player $1$ to deviate to $c_1$.
\end{example}

\section{A digression on continuous games}

In this section we highlight the complications discussed in Section \ref{sec:intro} by looking at how they manifest themselves in the geometry of Nash and correlated equilibria of continuous games\footnote{For concreteness' sake, define these to be games with finitely many players, compact metric strategy spaces, and continuous utility functions.}.  Existence of both types of equilibria in continuous games is generally proven in four steps: $1$) discretize the strategy spaces to obtain a finite game, $2$) prove the corresponding existence theorem for finite games, $3$) observe that an exact equilibrium of the finite game is an approximate equilibrium of the continuous game, and $4$) show that a limit of such approximate equilibria is an exact equilibrium of the continuous game.

Steps $1$ and $3$ are more or less content-free.  Step $2$ can be accomplished for Nash equilibria by fixed-point methods or for correlated equilibria by Hart and Schmeidler's clever minimax argument.  There is an interesting divergence at step $4$: in the case of Nash equilibria it is trivial and in the case of correlated equilibria it seems to require a hairy measure-theoretic argument (see \cite{hs:ece} or \cite{spo:cecgcc}).

The difference is easiest to understand in terms of an auxiliary map $\epsilon: \Delta(\Gamma)\to\Rm_{\geq 0}$.  This is defined by the condition that $\epsilon(\mu)$ is the smallest $\epsilon$ such that $\mu$ is an $\epsilon$-correlated equilibrium (no measurable deviation yields an expected gain of more than $\epsilon$).  Let $\epsilon^\Pi$ be the restriction of $\epsilon$ to $\Delta^\Pi(\Gamma)$.  Then $\epsilon^\Pi(\mu)$ is also the smallest $\epsilon$ such that $\mu$ is an $\epsilon$-Nash equilibrium in the usual sense.

We can study the map $\epsilon$ via two different topologies on its domain $\Delta(\Gamma)$: the weak and strong topologies.  Two measures are near each other in the strong topology if you can get from one to the other without moving much mass.  They are near each other in the weak topology if you can get from one to the other without moving much mass \emph{very far}.

It is easy to show that $\epsilon$ is continuous with respect to the strong topology, but this is not very useful because $\Delta(\Gamma)$ is not compact with respect to this topology (at least when $\Gamma$ is not a finite game) so the limiting arguments do not go through.  It is compact with respect to the weak topology and viewing $\Delta^\Pi(\Gamma)$ as a space of tuples it is easy to show that $\epsilon^\Pi$ is continuous with respect to the weak topology, hence we get the necessary limiting argument for the existence of Nash equilibria.

For correlated equilibria the situation is different: $\epsilon$ is not weakly continuous!  The argument is essentially the same as why $W^1$ was a correlated equilibrium in Example \ref{ex:correqjump} but $W^2$ was not.  We can move some mass a little bit (thinking now of strategically equivalent strategies as being quite ``close'' to each other) and change the information conveyed by the correlating device drastically.  This changes an exact correlated equilibrium to a distribution for which $\epsilon$ is bounded away from zero, no matter how small the distance we moved the mass.

In particular this shows that $\epsilon$ is not weakly upper semicontinuous.  In fact this doesn't matter; all we really need for the existence proof is weak lower semicontinuity.  It turns out $\epsilon$ has this property, but proving it is a delicate technical matter (again, see \cite{hs:ece} or \cite{spo:cecgcc}).

The fact that Nash equilibria behave nicely with respect to strategic equivalence is what allows us to turn problems regarding mixed Nash equilibria of polynomial games into finite-dimensional problems in moment spaces: see for example \cite{karlin:tig} and \cite{pp:polygames}.  The failure of correlated equilibria to behave nicely with respect to strategic equivalence is captured by the fact that the corresponding problems for correlated equilibria are inherently infinite-dimensional \cite{sop:sece}.

\section{A more specific problem}
The erroneous proof of Lemma $4.6$ claims to construct, for any symmetric mixed strategy $y$ of the minimizer in the game $(\compow{\Gamma})^0$, a mixed strategy $\pi^y\in\Delta_{G\times S_m}(\mpow{\Gamma})$ for the maximizer such that $u_M^0(\pi^y,y)=0$.  Furthermore, the construction implicitly claims that such a $\pi^y$ can be found which does not depend on the utilities of $\Gamma$ (as is the case in \cite{hs:ece}) and which is rational in $y$.  That is to say, the elements of $\pi^y$ are in the field generated by the elements of $y$ over $\Qm$.  For simplicity we will write $\pi^y\in\Qm[y]$.

Let us see how rationality of $\pi^y$ follows from the proof of Lemma $4.6$.  First note that a finite zero-sum game has a maximin strategy which is rational in the utilities, because the set of maximin strategies is a polytope defined by inequalities with coefficients linear in the utilities.  Therefore Lemmas $2.15$ and $3.8$ show that the certificates $\pi^y$ constructed in Hart and Schmeidler's proof of the existence of correlated equilibria and our proof of the existence of exchangeable equilibria are in $\Qm[y]$.  But Lemma $4.6$ claims to construct the certificate $\pi^y$ for order $m$ exchangeable equilibria by taking the one from Lemma $3.8$ for exchangeable equilibria and replacing it by the product of its marginals.  That is to say, we are taking the image of a tuple of elements of $\Qm[y]$ under a polynomial map, and so we end up with another tuple $\pi^y\in\Qm[y]$.

\begin{sidewaystable}
\centering
$\Gamma$
\vskip 12pt
\begin{tabular}{c||cc}
$(u_1,u_2)$ & $0$ & $1$ \\
\hline
\hline
$0$ & $(r,r)$ & $(s,0)$ \\
$1$ & $(0,s)$ & $(0,0)$
\end{tabular}
\vskip 36pt
$(\compowm{2}{\Gamma})^0$
\vskip 12pt
\scalebox{0.80}[0.80]{
\begin{tabular}{cccc || ccc | ccc | ccc | ccc | ccc | ccc | ccc | ccc}
\multicolumn{4}{r||}{Minimizer:}& \multicolumn{12}{c|}{Deviations for player $1$} & \multicolumn{12}{c}{Deviations for player $2$} \\
&&&& \multicolumn{12}{c|}{$\overbrace{\qquad\qquad\qquad\qquad\qquad\qquad\qquad\qquad\qquad\qquad\qquad\qquad\qquad\qquad\qquad}$} &\multicolumn{12}{c}{$\overbrace{\qquad\qquad\qquad\qquad\qquad\qquad\qquad\qquad\qquad\qquad\qquad\qquad\qquad\qquad\qquad}$} \\
 &&&& $00$ & $00$ & $00$ & $01$ & $01$ & $01$ & $10$ & $10$ & $10$ & $11$ & $11$ & $11$ & $00$ & $00$ & $00$ & $01$ & $01$ & $01$ & $10$ & $10$ & $10$ & $11$ & $11$ & $11$ \\
\multicolumn{4}{l||}{Maximizer:} & $\downarrow$ & $\downarrow$ & $\downarrow$ & $\downarrow$ & $\downarrow$ & $\downarrow$ & $\downarrow$ & $\downarrow$ & $\downarrow$ & $\downarrow$ & $\downarrow$ & $\downarrow$ & $\downarrow$ & $\downarrow$ & $\downarrow$ & $\downarrow$ & $\downarrow$ & $\downarrow$ & $\downarrow$ & $\downarrow$ & $\downarrow$ & $\downarrow$ & $\downarrow$ & $\downarrow$ \\
$s_2^2$ & $s_2^1$ & $s_1^2$ & $s_1^1$ & $01$ & $10$ & $11$ & $00$ & $10$ & $11$ & $00$ & $01$ & $11$ & $00$ & $01$ & $10$ &$01$ & $10$ & $11$ & $00$ & $10$ & $11$ & $00$ & $01$ & $11$ & $00$ & $01$ & $10$  \\
\hline
\hline
$0$ & $0$ & $0$ & $0$ & $r$ & $r$& $2r$ & $0$ & $0$ & $0$ & $0$ & $0$ & $0$ & $0$ & $0$ & $0$
& $r$ & $r$& $2r$ & $0$ & $0$ & $0$ & $0$ & $0$ & $0$ & $0$ & $0$ & $0$ \\
$0$ & $0$ & $0$ & $1$ & $0$ & $0$& $0$ & $-r$ & $0$ & $r$ & $0$ & $0$ & $0$ & $0$ & $0$ & $0$
& $s$ & $r$& $r+s$ & $0$ & $0$ & $0$ & $0$ & $0$ & $0$ & $0$ & $0$ & $0$ \\
$0$ & $0$ & $1$ & $0$ & $0$ & $0$& $0$ & $0$ & $0$ & $0$ & $-r$ & $0$ & $r$ & $0$ & $0$ & $0$
& $r$ & $s$& $r+s$ & $0$ & $0$ & $0$ & $0$ & $0$ & $0$ & $0$ & $0$ & $0$ \\
$0$ & $0$ & $1$ & $1$ & $0$ & $0$& $0$ & $0$ & $0$ & $0$ & $0$ & $0$ & $0$ & $-2r$ & $-r$ & $-r$
& $s$ & $s$& $2s$ & $0$ & $0$ & $0$ & $0$ & $0$ & $0$ & $0$ & $0$ & $0$ \\
\hline
$0$ & $1$ & $0$ & $0$ & $s$ & $r$& $r+s$ & $0$ & $0$ & $0$ & $0$ & $0$ & $0$ & $0$ & $0$ & $0$ & $0$ & $0$ & $0$ & $-r$ & $0$& $r$  & $0$ & $0$ & $0$ & $0$ & $0$ & $0$ \\
$0$ & $1$ & $0$ & $1$ & $0$ & $0$& $0$ & $-s$ & $r-s$ & $r$ & $0$ & $0$ & $0$ & $0$ & $0$ & $0$ & $0$ & $0$ & $0$ & $-s$ & $r-s$& $r$  & $0$ & $0$ & $0$ & $0$ & $0$ & $0$ \\
$0$ & $1$ & $1$ & $0$ & $0$ & $0$& $0$ & $0$ & $0$ & $0$ & $-r$ & $s-r$ & $s$ & $0$ & $0$ & $0$ & $0$ & $0$ & $0$ & $-r$ & $s-r$& $s$  & $0$ & $0$ & $0$ & $0$ & $0$ & $0$ \\
$0$ & $1$ & $1$ & $1$ & $0$ & $0$& $0$ & $0$ & $0$ & $0$ & $0$ & $0$ & $0$ & $-r-s$ & $-r$ & $-s$  & $0$ & $0$ & $0$ & $-s$ & $0$ & $s$ & $0$ & $0$ & $0$ & $0$ & $0$ & $0$ \\
\hline
$1$ & $0$ & $0$ & $0$ & $r$ & $s$& $r+s$ & $0$ & $0$ & $0$ & $0$ & $0$ & $0$ & $0$ & $0$ & $0$ & $0$ & $0$ & $0$  & $0$ & $0$ & $0$  & $-r$ & $0$& $r$ & $0$ & $0$ & $0$ \\
$1$ & $0$ & $0$ & $1$ & $0$ & $0$& $0$ & $-r$ & $s-r$ & $s$ & $0$ & $0$ & $0$ & $0$ & $0$ & $0$ & $0$ & $0$ & $0$  & $0$ & $0$ & $0$& $-r$ & $s-r$& $s$  & $0$ & $0$ & $0$ \\
$1$ & $0$ & $1$ & $0$ & $0$ & $0$& $0$ & $0$ & $0$ & $0$ & $-s$ & $r-s$ & $r$ & $0$ & $0$ & $0$ & $0$ & $0$ & $0$  & $0$ & $0$ & $0$ & $-s$ & $r-s$& $r$  & $0$ & $0$ & $0$ \\
$1$ & $0$ & $1$ & $1$ & $0$ & $0$& $0$ & $0$ & $0$ & $0$ & $0$ & $0$ & $0$ & $-r-s$ & $-s$ & $-r$  & $0$ & $0$ & $0$ & $0$ & $0$ & $0$ & $-s$ & $0$ & $s$  & $0$ & $0$ & $0$ \\
\hline
$1$ & $1$ & $0$ & $0$ & $s$ & $s$& $2s$ & $0$ & $0$ & $0$ & $0$ & $0$ & $0$ & $0$ & $0$ & $0$ & $0$ & $0$ & $0$  & $0$ & $0$ & $0$ & $0$ & $0$ & $0$ & $-2r$ & $-r$& $-r$  \\
$1$ & $1$ & $0$ & $1$ & $0$ & $0$& $0$ & $-s$ & $0$ & $s$ & $0$ & $0$ & $0$ & $0$ & $0$ & $0$ & $0$ & $0$ & $0$  & $0$ & $0$ & $0$  & $0$ & $0$ & $0$ & $-r-s$ & $-r$& $-s$  \\
$1$ & $1$ & $1$ & $0$ & $0$ & $0$& $0$ & $0$ & $0$ & $0$ & $-s$ & $0$ & $s$ & $0$ & $0$ & $0$ & $0$ & $0$ & $0$  & $0$ & $0$ & $0$  & $0$ & $0$ & $0$ & $-r-s$ & $-s$& $-r$ \\
$1$ & $1$ & $1$ & $1$ & $0$ & $0$& $0$ & $0$ & $0$ & $0$ & $0$ & $0$ & $0$ & $-2s$ & $-s$ & $-s$  & $0$ & $0$ & $0$ & $0$ & $0$ & $0$ & $0$ & $0$ & $0$ & $-2s$ & $-s$ & $-s$ \\
\end{tabular}}
\caption{Above: Utilities for $\Gamma$.  Player $1$ chooses rows and player $2$ chooses columns.  Below: Utilities for the maximizer in $(\compowm{2}{\Gamma})^0$.  The maximizer chooses rows and the minimizer chooses columns.  Columns associated with a deviation from a given strategy to itself would have all zeros and have been omitted.}
\label{tab:power2}
\end{sidewaystable}

\begin{example}In fact such a $\pi^y$ need not exist for order $m$ exchangeable equilibria, as the following example due to Sergiu Hart shows.  Consider the symmetric bimatrix game $\Gamma$ and its contracted second power $\compowm{2}{\Gamma}$ shown in Table \ref{tab:power2}.  Suppose the minimizer plays the mixed strategy $y_{\alpha,\beta}$ which assigns probability $\alpha$ to $00\rightarrow 01$ and $00\rightarrow 10$ and probability $\beta$ to $11\rightarrow 01$ and $11\rightarrow 10$ for each player, so we have $4(\alpha+\beta)=1$.

To have $\pi^{y_{\alpha,\beta}}\in\Delta_{\Zm_2\times S_2}(\mpowm{2}{\Gamma})$ means that $\pi^{y_{\alpha,\beta}}$ is a product of four i.i.d.\ $\operatorname{Bernoulli}(p)$ random variables for some $p\in [0,1]$.  For notational simplicity we will let $q=1-p$.  Then we can compute the expected utility of the maximizer in $(\compowm{2}{\Gamma})^0$ as:
\begin{equation*}
u_M^0(\pi^{y_{\alpha,\beta}},y_{\alpha,\beta}) = 4(\alpha q^2 - \beta p^2)(qr+ps).
\end{equation*}
For this expression to be zero for all $r$ and $s$ we must have $\alpha q^2 = \beta p^2$, so
\begin{equation*}
p = \frac{\sqrt{\alpha}}{\sqrt{\alpha}+\sqrt{\beta}}\qquad\text{and}\qquad q = \frac{\sqrt{\beta}}{\sqrt{\alpha}+\sqrt{\beta}},
\end{equation*}
and these expressions are clearly not rational in $\alpha$ and $\beta$.  But $p$ and $q$ are marginals of $\pi^{y_{\alpha,\beta}}$ and so can be written as sums of its elements.  Thus the entries of $\pi^{y_{\alpha,\beta}}$ cannot all be rational in $\alpha$ and $\beta$ either.
\end{example}

\bibliographystyle{plain}
\bibliography{../../references}